\documentclass[sigconf]{acmart}

\renewcommand\footnotetextcopyrightpermission[1]{}

\usepackage{verbatim}
\usepackage{xcolor}
\usepackage[normalem]{ulem}

\usepackage{listings}

\definecolor{codegreen}{rgb}{0,0.6,0}
\definecolor{codegray}{rgb}{0.5,0.5,0.5}
\definecolor{codepurple}{rgb}{0.58,0,0.82}
\definecolor{backcolour}{rgb}{0.95,0.95,0.92}

\lstdefinestyle{mystyle}{
	commentstyle=\color{codegreen},
	keywordstyle=\color{magenta},
	numberstyle=\tiny\color{codegray},
	stringstyle=\color{codepurple},
	basicstyle=\ttfamily\footnotesize,
	breakatwhitespace=false,         
	breaklines=true,                 
	captionpos=b,                    
	keepspaces=true,                 
	numbers=left,                    
	numbersep=5pt,                  
	showspaces=false,                
	showstringspaces=false,
	showtabs=false,                  
	tabsize=2,
	frame=single
}

\usepackage{graphicx}

\usepackage{tikz}
\usetikzlibrary{decorations.pathreplacing}
\usetikzlibrary{positioning}
\usetikzlibrary{fit}

\usepackage{hyperref}

\usepackage{pgfplots}
\pgfplotsset{width=\columnwidth,compat=1.9}
\usepackage{caption}
\usepackage{subcaption}

\newcommand{\uop}{{$\mu$op}}
\newcommand{\uops}{{\uop s}}
\newcommand{\offset}{{offset}}
\newcommand{\obsA}{\texttt{noMFuse}}
\newcommand{\obsB}{\texttt{no$\mu$Cache}}

\usepackage{amsthm}
\theoremstyle{definition}
\newtheorem*{theorem*}{\textit{Hypothesis}}

\usepackage{enumitem}

\usepackage{hyperref}
\usepackage[capitalise]{cleveref}

\usepackage{fancyvrb}

\usepackage{array}
\newcolumntype{H}{>{\setbox0=\hbox\bgroup}c<{\egroup}@{}}
\newcolumntype{P}[1]{>{\centering\arraybackslash}p{#1}}
\newcolumntype{?}{!{\vrule width 1pt}}
\usepackage{booktabs}
\usepackage{multirow}
\usepackage{color, colortbl}

\usepackage{balance}
\usepackage{stfloats}

\begin{document}

\title{On Abnormal Execution Timing of Conditional Jump Instructions}

\author{Annika Wilde}
\email{annika.wilde@rub.de}
\affiliation{%
  \institution{Ruhr University Bochum}
  \country{Germany}}

\author{Samira Briongos}
\email{samirabriongos@gmail.com}
\affiliation{%
  \institution{NEC Laboratories Europe}
  \country{Germany}}

\author{Claudio Soriente}
\email{csoriente@gmv.com}
\affiliation{%
  \institution{GMV Spain}
  \country{Spain}}

\author{Ghassan Karame}
\email{ghassan@karame.org}
\affiliation{%
  \institution{Ruhr University Bochum}
  \country{Germany}}

\begin{abstract}
	
    An extensive line of work on modern computing architectures has shown that the execution time of instructions can (i) depend on the operand of the instruction or (ii) be influenced by system optimizations, e.g., branch prediction and speculative execution paradigms.

    In this paper, we systematically measure and analyze timing variabilities in conditional jump instructions that can be macro-fused with a preceding instruction, depending on their placement within the binary. Our measurements indicate that these timing variations stem from the \uop{} cache placement and the jump's offset in the L1 instruction cache of modern processors. We demonstrate that this behavior is consistent across multiple microarchitectures, including Skylake, Coffee Lake, and Kaby Lake, as well as various real-world implementations. 
    We confirm the prevalence of this variability through extensive experiments on a large-scale set of popular binaries, including libraries from Ubuntu 24.04, Windows 10 Pro, and several open-source cryptographic libraries. We also show that one can easily avoid this timing variability by ensuring that macro-fusible instructions are 32-byte aligned---an approach initially suggested in 2019 by Intel in an overlooked short report. We quantify the performance impact of this approach across the cryptographic libraries, showing a speedup of 2.15\% on average (and up to 10.54\%) when avoiding the timing variability.
   As a by-product, we show that this variability can be exploited as a covert channel, achieving a maximum throughput of 16.14 Mbps.
    
\end{abstract}

\maketitle
\pagestyle{plain}

\section{Introduction}

Software instructions exhibit unintended variations in their execution time that may affect a program's performance and security.
For example, Großschädl \emph{et al.}~\cite{grossschaedl2010scEarlyTerminationMultiplication} and Andrysco \emph{et al.}~\cite{andrysco2015subnormalFloatingPointTiming} demonstrated that the type of operand(s) influences the execution time of multiplications and floating-point operations on certain processors. 
Other works have uncovered and exploited time variations arising from cache memories~ \cite{yarom2014flushReload, almeida2016verifyingConstantTime, purnal2021primeScope}, port contention~\cite{aldaya2019portContention}, speculative execution~\cite{canella2020meltdown,kocher2018spectreAttacksExploitingSpeculative}, and branch predictions~\cite{aciicmez2006predictingSecretKeysBranchPrediction}.
Another line of work has explored timing fluctuations caused by the decode procedure in the processor's front-end. 
For example, Kim \emph{et al.}~\cite{kim2021uccheckUopCachex86} and Ren~\emph{et~al.}~\cite{ren2021deadUops} reverse engineer the \uop{} cache and show that cache contention results in measurable timing differences that can be exploited to exfiltrate information. Deng \emph{et al.}~\cite{deng2022leakyFrontends} characterize and exploit diverging decode times of the available decode paths. 
Finally, Wang \emph{et al.}~\cite{wang2023crossCoreMacroOpFusion} analyze stalls in the decoder and measure their security impact. 

In this paper, we take a different approach and systematically analyze and measure the timing variations of conditional jump instructions due to instruction placement in the binary. 
In particular, our measurements show that the position of a conditional jump in the binary affects macro-op fusion and the caching of the corresponding \uops{} in the \uop{} cache, which in turn determines the instruction fetch time. We stress that such behavior (i) does not depend on the instruction operand (e.g., the jump target), (ii) is orthogonal to both branch prediction and speculative execution, and (iii) does not rely on cache contention. 

We measure the prevalence of this timing variability by means of thorough experiments and show that up to 15.6\% of all macro-fusible conditional jump instructions identified in our experiments are likely to be on a ``slow path''---incurring additional cycles for execution. Among these conditional jumps that require additional cycles, our findings suggest that 10\% of those result from failed macro-op fusion, while the remaining 90\% are caused by a microcode update introduced by Intel, resulting in an increased number of fetches due to \uop{} cache misses (cf. Section~\ref{sec:problem}). 
We demonstrate that this behavior is consistent across multiple microarchitectures, including Skylake, Coffee Lake, and Kaby Lake. Surprisingly, our measurements show that AMD architectures exhibit similar behavior, though driven by different factors; namely, we show that 7.8\% of all jump instructions in AMD Zen 3 are likely to exhibit a slow path due to an increased number of micro-op cache fetches (cf. \Cref{sec:reproducibility}). 

We validate our measurements with a large-scale study of 666 real-world libraries, comprising 320 from a fresh Ubuntu 24.04.2 installation, 323 from a fresh Windows 10 Pro 22H2 installation, and 23 open-source cryptographic libraries. Surprisingly, the empirical results of our static analysis show that 99.7\% of the Ubuntu libraries, 87.0\% of the Windows libraries, and all cryptographic libraries conform to the expected 16\% rate of macro-fusible conditional jumps on the slow path (cf. \Cref{sec:performance}).
These findings suggest that there is additional room to speed up the execution time. Namely, practitioners can easily reduce the execution time of their code by ensuring that macro-fusible instructions do not span 32-byte boundaries---an approach initially suggested in 2019 by Intel in an overlooked short report~\cite{intel2024mitigationJccErratum}. We quantify the performance impact of this approach by applying it to the set of cryptographic libraries. Namely, our experimental results show that this countermeasure yields an average improvement of 2.15\%---with gains of up to 10.5\% in some open-source cryptographic libraries (cf. \Cref{sec:mitigation_significance}).

As a by-product, we note that these timing variabilities in conditional jumps can allow an adversary to distinguish which branches are being executed in software---even when all branches are balanced. We show, using empirical analysis, that an adversary can achieve a maximum throughput of up to 16.14 Mbps by exploiting such variabilities to create a cross-core covert channel between two processes  (Section~\ref{sec:attack}).

We are currently in the process of responsibly disclosing our findings to Ubuntu, Microsoft, and the developers of the analyzed cryptographic libraries. 
At the time of writing, both WolfSSL and Amazon s2n-tls have officially adopted our recommendations.

\section{Background \& Related Work}

In this section, we provide an overview of the x86 microarchitecture and the organization of the \uop{} cache, and briefly discuss related work on timing variability in the x86 front-end.

\subsection{x86 Microarchitecture}
\label{sec:uarch}

\begin{figure}
    \centering
    \includegraphics[width=1.0\linewidth]{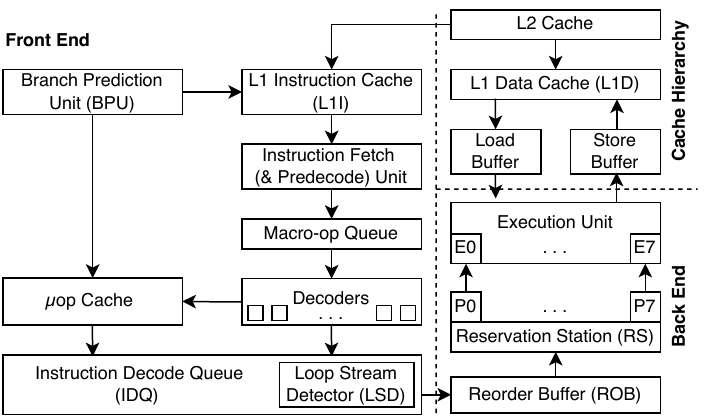}
    \caption{Overview of the x86 microarchitecture.}
    \label{fig:uarch}
\end{figure}

The x86 microarchitecture typically comprises three high-level components: the front-end, the back-end, and the memory subsystem. These components handle various tasks, such as fetching, decoding, executing instructions, and retrieving data. Figure~\ref{fig:uarch} illustrates the critical elements of the x86 microarchitecture and their relationships.

\vspace{0.5 em}\noindent \textbf{Cache Hierarchy.} 
Memory is organized in pages, typically of 4KB size. Given that access to the main memory incurs a high latency, modern processors feature three levels of cache: Last Level Cache (LLC), Level 2 Cache (L2), and Level 1 Cache (L1). The L1 cache has the least latency and is, in contrast to the other caches, divided into an instruction (L1I) and data cache (L1D). Recent x86 CPUs feature an additional micro-op cache (\uop{} cache), which stores the translation of instructions into micro-ops and has an even lower latency than the L1I cache (cf. Section~\ref{sec:uop_cache}).   

Note that the L2 and L1D caches belong to the memory subsystem, whereas the L1I cache belongs to the front-end.
Also, L1D is accessed by the back-end through the load and store buffers which accelerate the memory accesses.

\vspace{0.5 em}\noindent \textbf{The front-end. } 
The x86 instruction set architecture (ISA) defines a set of complex instructions with varying instruction lengths. 
The front-end is responsible for fetching and decoding these ``variable-length'' instructions. Memory lines are loaded into the L1I cache from the L2 cache. Current x86 processors can fetch up to 64 bytes per cycle. The \emph{instruction fetch unit} then fetches 16- or 32-byte–aligned blocks from the L1I cache. In Intel processors, the fetch unit includes a \emph{pre-decoder} that marks the bounds of the variable-length instructions for subsequent processing. The pre-decoder outputs so-called \emph{macro-ops}, which are stored in the \emph{macro-op queue}. Note that the instruction fetch unit only forwards complete macro-ops. 
{When an instruction crosses a fetch boundary, the unit waits until the following cycle, after the rest of the instruction bytes are fetched, before forwarding the complete macro-op.}

The macro-op queue buffers macro-ops and optionally performs \emph{macro-op fusion}, where specific logic, arithmetic, and compare instructions are fused with a subsequent jump instruction to a single macro-op performing both operations. 

On the other hand, the decoding unit translates the macro-ops into \emph{micro-ops (\uops)}. Unlike macro-ops, \uops{} are fixed in length and exhibit reduced complexity, enabling more efficient processing within the microarchitecture. The number of decoders in the decoding unit and the amount of \uops{} output per cycle vary among manufacturers and architectures. For instance, the decoding unit of Intel's Skylake microarchitecture contains five decoders and outputs up to five \uops{} per cycle. The decoded \uops{} are stored in the \emph{\uop{} cache} to bypass the decoding for previously decoded instructions. The \emph{branch prediction unit (BPU)} tries to predict whether a branch will be taken and, consequently, determines the address of the next instruction. Based on this prediction, the BPU checks the corresponding address in the \uop{} cache. Hence, \uops{} are either provided by the \uop{} cache (\uop{} cache hit) or by the decoder (\uop{} cache miss), with the \uop{} cache offering higher throughput. 

Finally, the \uops{} are stored in the \emph{instruction decode queue (IDQ)}, which buffers between the front-end and back-end to mitigate stalls. Front-end stalls can arise from BPU mispredictions, \uop{} cache misses that require refilling the decode pipeline, or L1I cache misses. Insufficient \uop{} delivery may empty the IDQ, causing stalls due to the lack of \uops{} for dispatch. Conversely, back-end stalls arise when it temporarily exceeds its capacity to process \uops{}, prompting the IDQ to pause delivery. In some processors, the IDQ incorporates a loop stream detector (LSD), which detects loops up to 70 \uops{} and streams them directly.

\begin{figure}
	\centering
	\includegraphics[width=1.0\linewidth]{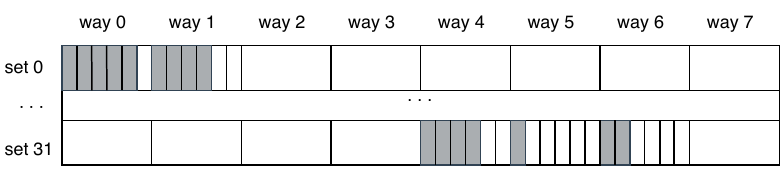}
	\caption{Overview of the micro-op cache of Intel's Skylake microarchitecture. Gray fields represent used \uop{} slots, indicating that the cache holds valid \uops{} ready to be streamed to the IDQ. White fields represent empty or skipped slots due to the cache’s placement policy or alignment restrictions.
	}
	\label{fig:uopcache}
\end{figure}

\vspace{0.5 em}\noindent \textbf{The back-end.}
The back-end executes \uops{}. Namely, it receives a stream of in-order \uops{} from the IDQ and stores them in the \emph{reorder buffer (ROB)}. 
The ROB performs optimizations, such as register renaming, elimination of register-to-register moves, and zeroing idioms. These optimizations allow skipping the execution of certain \uops{} or to parallelize them by breaking dependencies. 
The ROB then delivers the \uops{} to the \emph{reservation station (RS)}, which schedules them to suitable ports for delivery to the respective execution units. 
Each execution unit serves a specific function, such as arithmetic computation. Some units are duplicated to increase throughput and enhance parallel processing capabilities. At execution time, sequential \uops{} may be executed in parallel or even out-of-order if they are independent. The ROB then sequentially retires the \uops{} in the correct order as received from the IDQ.

\begin{figure*}[tbp]
	\centering
	\begin{subfigure}[t]{0.19\textwidth}
		\centering
		\includegraphics[width=1.0\linewidth]{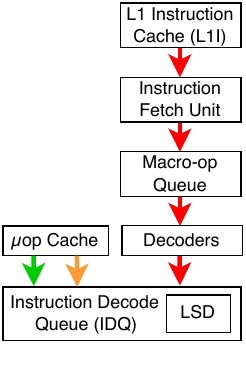}
		\vspace{-0.8cm}
		\caption{\cite{ren2021deadUops,kim2021uccheckUopCachex86}}
		\label{fig:rel_dead_uops}
	\end{subfigure}
	\hfill
	\begin{subfigure}[t]{0.19\textwidth}
		\centering
		\includegraphics[width=1.0\linewidth]{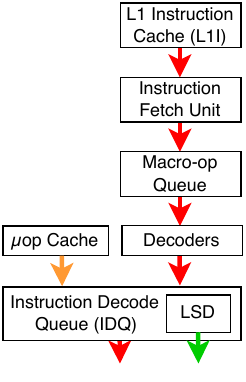}
		\vspace{-0.8cm}
		\caption{\cite{deng2022leakyFrontends}}
		\label{fig:rel_leaky_fe}
	\end{subfigure}
	\hfill
	\begin{subfigure}[t]{0.19\textwidth}
		\centering
		\includegraphics[width=1.0\linewidth]{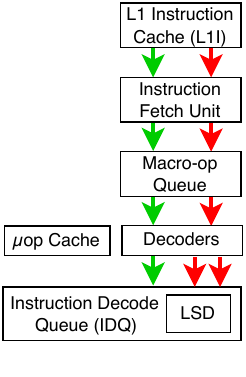}
		\vspace{-0.8cm}
		\caption{\cite{wang2023crossCoreMacroOpFusion}}
		\label{fig:rel_mop_fusion}
	\end{subfigure}
	\unskip\ \vrule\ 
	\begin{subfigure}[t]{0.19\textwidth}
		\centering
		\includegraphics[width=1.0\linewidth]{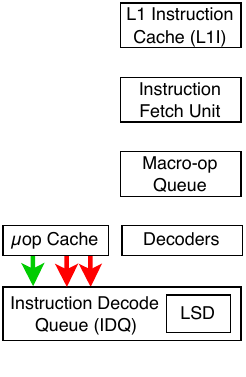}
		\vspace{-0.8cm}
		\caption{\obsA}
		\label{fig:our_mop_fusion}
	\end{subfigure}
	\hfill
	\begin{subfigure}[t]{0.19\textwidth}
		\centering
		\includegraphics[width=1.0\linewidth]{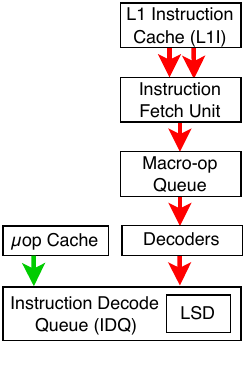}
		\vspace{-0.8cm}
		\caption{\obsB}
		\label{fig:our_misalignment}
	\end{subfigure}
	\caption{Overview of the x86 fetch paths in the front-end explored in related work (a-c) and this paper (d-e). Green arrows indicate the fast path, whereas red arrows denote the slow path. Two red arrows on the same path denote an additional cycle required for fetching. An orange arrow from the \uop{} cache represents a scenario in which the IDQ receives some \uops{} from the \uop{} cache while others originate from the decoding pipeline, introducing delays for refilling the decoding pipeline.
	}
	\label{fig:problem}
\end{figure*}

\subsection{Micro-op Cache}
\label{sec:uop_cache}

The micro-op cache buffers translated \uops{}, allowing to bypass the decode pipeline (i.e., fetching, pre-decoding, and decoding) for recently decoded \uops. The \uop{} cache is organized in cache sets and cache lines. Figure~\ref{fig:uopcache} illustrates the \uop{} cache of Intel's Skylake architecture. Here, the \uop{} cache consists of 32 sets and eight ways, and can hold up to six \uops{} per cache line (i.e., a specific set and way)~\cite{agner2024microarchitecture}. The cache set is indexed by the bits 5–9 of the respective instruction. It differs from other cache levels in that the \uop{} cache has complex additional placement rules and does not buffer every \uop{}~\cite{ren2021deadUops}. For instance, the \uop{} cache can only store \uops{} of a 32-byte-aligned code region if the corresponding \uops{} fit within three cache lines. Otherwise, the region will always be fetched from the decode pipeline. Additionally, \uops{} derived from the same macro-op must be placed in the same cache line. Finally, an unconditional jump must always be the last \uop{} in a cache line, among other placement rules. Hence, a \uop{} cache line may have empty slots depending on these constraints, as illustrated in Figure~\ref{fig:uopcache}. The \uop{} cache of AMD CPUs is organized similarly with deviating sizes~\cite{kim2021uccheckUopCachex86}.

The \uop{} cache can stream \uops{} corresponding to up to 64B per cycle in some processors, significantly increasing the throughput. The decode pipeline is inactive while \uops{} are streamed from the \uop{} cache. If a \uop{} cache miss occurs, the front-end switches to decode the pipeline, incurring a delay to refill the pipeline.

\subsection{Related Work}
\label{sec:related_frontend}

Previous research has highlighted the time variations due to the different paths that a code block can take to reach the back-end. 
Ren \emph{et al.}~\cite{ren2021deadUops} and Kim \emph{et al.}~\cite{kim2021uccheckUopCachex86} 
focus on reverse engineering the \uop{} cache and leverage these insights to create covert- and side-channels based on \uop{} cache contention.
Deng \emph{et al.}~\cite{deng2022leakyFrontends} explore the timing variations due to the different paths a block of code can take in the decode pipeline; they show a covert-channel where the sender forces specific decode paths to transmit information to a receiver. Finally, Wang \emph{et al.}~\cite{wang2023crossCoreMacroOpFusion} explore stalls in the decoding unit triggered by specific combinations of fusible macro-ops. Decoders have a limited capacity for macro-op fusion per cycle, and once this limit is reached, decoding halts. This results in an additional cycle to process the same number of instructions, leading to a measurable delay.

Figure~\ref{fig:problem}(a-c) offers a detailed overview of the front-end timing variations studied in previous research. We point out that none of these works~\cite{ren2021deadUops,kim2021uccheckUopCachex86,deng2022leakyFrontends,wang2023crossCoreMacroOpFusion} covers the front-end timing \linebreak abnormalities that we discuss in Section~\ref{sec:problem} or measures the performance impact of the timing variations in real-world binaries. 
Furthermore, the covert-channels introduced  in~\cite{ren2021deadUops,kim2021uccheckUopCachex86,deng2022leakyFrontends,wang2023crossCoreMacroOpFusion} rely on either cache contention or stalls in the front-end decoder; in contrast, the covert-channel that we describe in Section~\ref{sec:attack} does not require any contention on shared resources between the sender and the receiver.

\section{Abnormal Timing in Conditional Jumps}
\label{sec:problem}

As mentioned in Section~\ref{sec:related_frontend}, execution-time variability due to the front-end of x86 microarchitectures has been thoroughly analyzed by~\cite{deng2022leakyFrontends,ren2021deadUops,kim2021uccheckUopCachex86,wang2023crossCoreMacroOpFusion}. 
 
In this paper, we systematically measure and analyze two additional behaviors of x86 front-ends that incur timing differences in conditional jumps based on the instruction placement within the binary.

\subsection{\obsA : Impact of Misalignment on Macro-Op Fusion}
\label{sec:root_cause_A}

Modern x86 microarchitectures optimize performance by fusing certain logic, arithmetic, and comparison instructions with a subsequent jump instruction into a single macro-op, which is then translated into a single \uop. This optimization reduces the load on the \uop{} cache and downstream microarchitectural components. However, we observe that specific placement of jump instructions can prevent macro-op fusion~\cite{intel2024optimizationManual}. In such cases, the two instructions remain unfused and are translated into separate \uops{}. Additionally, this placement causes the resulting \uops{} to be stored in distinct \uop{} cache sets. Since the IDQ cannot fetch \uops{} from two different \uop{} cache sets within the same cycle, two cycles are required to fetch both instructions (cf. Figure~\ref{fig:our_mop_fusion}). In contrast, when the instructions are macro-fused and translated into a single \uop{}, they can be fetched in a single cycle. This difference introduces a {one-cycle} timing variation, depending on the placement of the fusible instruction pair.

\begin{figure}
    \centering
    \includegraphics[width=0.58\linewidth]{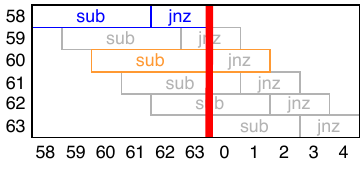}
    \caption{Different placement of a macro-fusible \texttt{sub} and \texttt{jnz} instruction within a 64-byte block. The 32/64B-aligned boundary is marked in red. Those instructions not crossing this boundary are marked in blue, those preventing macro-op fusion (\obsA) are highlighted in orange, and the other ones (\obsB) are marked in gray.}
    \label{fig:instruction_shift}
\end{figure}

We identify three different types of instruction placement that influence fetching. Figure~\ref{fig:instruction_shift} illustrates these placement scenarios using a four-byte \texttt{sub} instruction followed by a two-byte \texttt{jnz} instruction. In the first case, highlighted in blue, both instructions are located within the same 32-byte block, ensuring that they are consistently macro-fused and can be fetched within a single cycle from either the decode pipeline or the \uop{} cache. However, when the instruction pair spans a 32-byte boundary, an additional cycle is required for fetching, primarily due to delays in the instruction fetch unit. A special case arises when the instructions \texttt{sub} and \texttt{jnz} are placed in separate 64-byte blocks (highlighted in orange). This specific placement prevents macro-op fusion, as we further investigate in this section. Based on this analysis, a jump instruction encounters this special case with a probability of $\frac{1}{64}=1.56\%$; that is, only one out of the 64 possible jump alignments within a 64-byte cache line {positions the jump} precisely at the boundary. In this scenario, the preceding fusible instruction lacks sufficient space within the same cache line, thereby preventing macro-op fusion.

In what follows, we additionally measure the impact of macro-op fusion on the front-end (which we dub \obsA) and analyze the timing exhibited by an exemplary loop (cf. Figure~\ref{lst:instrumented_loop}). 
Notice that the loop contains only two instructions: a \texttt{sub} instruction, decrementing the loop counter at each iteration (cf. line 13), and a \texttt{jnz} instruction, jumping back to the beginning of the loop if the counter is not zero (cf. line 14).

\vspace{0.5 em}\noindent \textbf{Setup. }
Unless stated otherwise, we execute all experiments on an Intel Xeon E-2286G CPU, which is based on the Coffee Lake microarchitecture, with 128GB RAM running Ubuntu 20.04 LTS.
We reserve a physical core for the experiments using the \texttt{isolcpus} option at boot-up and disable hyperthreading, ensuring that neither hyperthreading nor other processes interfere with the experiments.
We then enforce the execution of the experiments on the reserved core.
In addition, we fix the CPU frequency to disable dynamic voltage and frequency scaling (DVFS).
Last but not least, we flush all caches before starting each measurement.
We also repeat the experiments over multiple days after restarting the machine to ensure that a primed machine state does not influence the results.

Our experiments are based on the code snippet in Figure~\ref{lst:instrumented_loop}.
Lines 1-3 implement an optional padding with \texttt{NOP} instructions to change the location of the loop in the compiled binary.
We repeat the padding instruction \texttt{B} times to achieve a shift by \texttt{B} bytes.
We vary \texttt{B} between zero and 63 to encompass all possible offsets within an L1I cache line.
We measure the timing difference with the \texttt{rdtsc} instruction in lines 8 and 17, which returns the number of cycles since the last restart. 
The counter provides an accuracy of 1-3 cycles on Intel CPUs.
We place fences before and after the \texttt{rdtsc} instruction to ensure that the previous code (e.g., the varying shift) does not interfere with our measurements.
We also monitor selected hardware performance counters (HPCs) to examine the behavior of the respective micro-architectural components in lines 6 and 19. Table~\ref{tab:hpc_list} summarizes the observed HPCs and the specific events they capture.
Line 11 initializes a counter to determine the number of consecutive jump instructions executed during a single measurement.
We set the counter to 200, ensuring that the loop executes 200 jumps (to achieve a stable output execution time for profiling).
Line 13 then decrements the counter and sets the flags for the jump instruction in line 14. 
All data points are averaged over 1000 independent runs.

\begin{figure}[tb]
    \centering
    \scriptsize
\begin{Verbatim}[commandchars=\\\{\},frame=single,numbers=left,xleftmargin=5mm]
\textcolor{blue}{.rept B}  
    \textcolor{blue}{NOP} # shift the loop position
\textcolor{blue}{.endr}

lfence 
call PAPI_read
lfence 
rdtsc
lfence

mov rcx,0xc8
loop:
    sub rcx, 1
    jnz loop

lfence 
rdtsc
lfence 
call PAPI_read
\end{Verbatim}
    \caption{Example of a loop that decrements a counter by one and jumps to the loop start if the counter is not zero. The loop is instrumented to measure the execution time and different hardware events. The position of the loop in the binary is changed by modifying the number \texttt{B} of \texttt{NOP} instructions.}
    \label{lst:instrumented_loop}
\end{figure}

\vspace{0.5 em}\noindent \textbf{Detailed findings.}
The overall timing variability of the code in Figure~\ref{lst:instrumented_loop} is depicted in Figure~\ref{fig:results_rdtsc}. Here, we define the \offset{} as the six least significant bits of the \texttt{loop} address. Our initial observation is based on execution timing: when placed in \offset s 27-31 or 59-63, the execution time increases by approximately 200 cycles compared to other \offset s. This corresponds to an overhead of approximately one cycle per iteration, averaged over 200 loop iterations. 

Moreover, we observe that the x86 microarchitecture processes loop instructions differently when placed at \offset{}  60. Table~\ref{tab:exceprt_obs1} pre-sents our measurement results for selected \offset s and HPCs to validate this observation. For the sake of simplicity, we discuss our results based on these representative \offset s and refer the reader to Figure~\ref{fig:results_all}(a-d) for a comprehensive view of all 64 \offset s.  
The first row of the table shows the execution time for \offset s 0, 32, and 60. Notably, \offset{} 60 exhibits an increased execution time of approximately 190 cycles compared to \offset s 0 and 32, confirming our initial observation. Note that this increased execution time is independent of branch prediction, as evidenced by the \texttt{BACLEARS.ANY} HPC, which reports a constant number of front-end re-steers. Furthermore, this behavior is not due to a difference in the number of executed instructions (see the third row of Table~\ref{tab:exceprt_obs1}).

\definecolor{clr1}{RGB}{245,139,62} 
\definecolor{clr2}{RGB}{64,147,197} 
\definecolor{clr3}{RGB}{203,204,204} 

\begin{figure}
    \centering
    \scalebox{0.9}{\begin{tikzpicture}
        \begin{axis}[
            x tick label style={
                /pgf/number format/1000 sep=},
            ylabel=\#cycles,
            xlabel=\offset ,
            enlargelimits=0.05,
            legend style={at={(0.5,-0.1)},
            anchor=north,legend columns=-1},
            bar width=2.5pt,
            ybar stacked,
            grid=none,
            xtick style={draw=none},
            ytick style={draw=none},
            xtick pos=left,
            height=4cm,
            ymajorgrids,
        ]
        
        \addplot[clr3!90,fill=clr3!90] 
            coordinates { (0, 304.6) (1, 318.6) (2, 323.3) (3, 324.2) (4, 311.9) (5, 308.9) (6, 307.3) (7, 324.5) (8, 307.0) (9, 304.8) (10, 313.0) (11, 362.2) (12, 354.8) (13, 358.5) (14, 363.7) (15, 347.1) (16, 307.1) (17, 307.4) (18, 307.0) (19, 307.2) (20, 309.0) (21, 311.8) (22, 305.7) (23, 308.8) (24, 325.8) (25, 328.4) (26, 322.2) (27, 0.0) (28, 0.0) (29, 0.0) (30, 0.0) (31, 0.0) (32, 307.3) (33, 327.5) (34, 309.2) (35, 322.0) (36, 304.6) (37, 310.6) (38, 345.4) (39, 307.0) (40, 328.5) (41, 316.0) (42, 322.0) (43, 358.0) (44, 354.4) (45, 343.9) (46, 356.3) (47, 354.2) (48, 307.9) (49, 308.6) (50, 317.8) (51, 307.8) (52, 311.1) (53, 305.1) (54, 313.0) (55, 311.4) (56, 306.3) (57, 313.4) (58, 331.8) (59, 0.0) (60, 0.0) (61, 0.0) (62, 0.0) (63, 0.0) };

        \addplot[clr2!90,fill=clr2!90] 
            coordinates { (0, 0.0) (1, 0.0) (2, 0.0) (3, 0.0) (4, 0.0) (5, 0.0) (6, 0.0) (7, 0.0) (8, 0.0) (9, 0.0) (10, 0.0) (11, 0.0) (12, 0.0) (13, 0.0) (14, 0.0) (15, 0.0) (16, 0.0) (17, 0.0) (18, 0.0) (19, 0.0) (20, 0.0) (21, 0.0) (22, 0.0) (23, 0.0) (24, 0.0) (25, 0.0) (26, 0.0) (27, 0.0) (28, 0.0) (29, 0.0) (30, 0.0) (31, 0.0) (32, 0.0) (33, 0.0) (34, 0.0) (35, 0.0) (36, 0.0) (37, 0.0) (38, 0.0) (39, 0.0) (40, 0.0) (41, 0.0) (42, 0.0) (43, 0.0) (44, 0.0) (45, 0.0) (46, 0.0) (47, 0.0) (48, 0.0) (49, 0.0) (50, 0.0) (51, 0.0) (52, 0.0) (53, 0.0) (54, 0.0) (55, 0.0) (56, 0.0) (57, 0.0) (58, 0.0) (59, 0.0) (60, 494.8) (61, 0.0) (62, 0.0) (63, 0.0)  };

        \addplot[clr1!90,fill=clr1!90] 
            coordinates { (0, 0.0) (1, 0.0) (2, 0.0) (3, 0.0) (4, 0.0) (5, 0.0) (6, 0.0) (7, 0.0) (8, 0.0) (9, 0.0) (10, 0.0) (11, 0.0) (12, 0.0) (13, 0.0) (14, 0.0) (15, 0.0) (16, 0.0) (17, 0.0) (18, 0.0) (19, 0.0) (20, 0.0) (21, 0.0) (22, 0.0) (23, 0.0) (24, 0.0) (25, 0.0) (26, 0.0) (27, 493.8) (28, 493.7) (29, 490.3) (30, 516.0) (31, 508.2) (32, 0.0) (33, 0.0) (34, 0.0) (35, 0.0) (36, 0.0) (37, 0.0) (38, 0.0) (39, 0.0) (40, 0.0) (41, 0.0) (42, 0.0) (43, 0.0) (44, 0.0) (45, 0.0) (46, 0.0) (47, 0.0) (48, 0.0) (49, 0.0) (50, 0.0) (51, 0.0) (52, 0.0) (53, 0.0) (54, 0.0) (55, 0.0) (56, 0.0) (57, 0.0) (58, 0.0) (59, 508.1) (60, 0.0) (61, 490.9) (62, 500.4) (63, 515.4)  };
        \end{axis}
    \end{tikzpicture}} 
    \caption{Execution time of the loop in Figure~\ref{lst:instrumented_loop} for varying \offset s. Offsets corresponding to \obsA{} are highlighted in blue, and those corresponding to \obsB{} are highlighted in orange. All other \offset s are represented in grey. Each data point is averaged over 1000 independent runs.}
    \label{fig:results_rdtsc}
\end{figure}
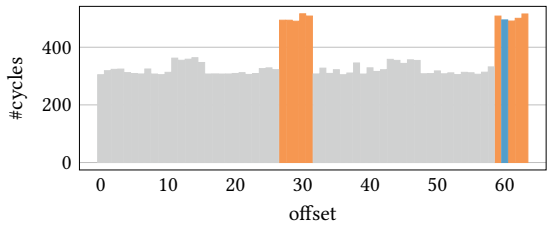

Instead, as shown in the fourth row, the primary cause of this discrepancy is a higher number of \uops{} issued. Specifically, the number of issued \uops{} for \offset s 0 and 32 remains around 600, whereas for \offset{} 60, the count approaches 800. This discrepancy arises due to an architectural feature of macro-op fusion in x86 processors, where the \texttt{sub} and \texttt{jnz} instructions can be fused to a single macro-op by the macro-op queue. The instruction decoders then translate this macro-fused instruction into a single \uop~\cite{wang2023crossCoreMacroOpFusion}, which occupies a single slot in the \uop{} cache.
However, Intel's x86 processors impose a restriction on macro-op fusion: if two fusible macro-ops are entirely located in different L1I cache lines, fusion does not occur~\cite{intel2024optimizationManual}. Specifically, fusion is prevented if the jump instruction is 64B aligned. In our example, this happens when the loop is placed at \offset{} 60. In this case, the two resulting \uops{} are placed in separate 32B aligned code blocks, causing them to be stored in separate sets within the \uop{} cache. 
Since two sets cannot be accessed simultaneously, an additional cycle is required to fetch the two \uops{} compared to the single \uop{} generated through macro-op fusion. Thus, the increased execution time that we measure at \offset{} 60 is a direct consequence of failed macro-op fusion due to L1I cache alignment constraints.

\begin{table}[tbp]
    \centering
    \caption{Events for a selected subset of \offset s and HPCs relevant for \obsA{} in the Coffee Lake microarchitecture.}
    \label{tab:exceprt_obs1}
    \scalebox{0.83}{\begin{tabular}{|l|l|c|c|c|} 
         \hline
         \multirow{2}{*}{\textbf{HPC}} & \multirow{2}{*}{\textbf{event}} & \multicolumn{3}{c|}{\textbf{\offset}} \\ \cline{3-5}
          & & \textbf{0} & \textbf{32} & \textbf{60}\\
         \hline
         \texttt{rdtsc}             & execution cycles      & 305 & 307 & 495 \\ \hline
         \texttt{BACLEARS.ANY}      & branch mispredictions &  29 &  29 &  28 \\ \hline
         \texttt{INSTRUCTIONS}      & executed instructions & 651 & 651 & 651 \\ \hline
         \texttt{UOPS\_ISSUED.ALL}  & issued \uops          & 611 & 616 & 780 \\ \hline
    \end{tabular}}
\end{table}

\subsection{\obsB: Impact of Misalignment on \uop-Caching}
\label{sec:root_cause_B}

In Section~\ref{sec:root_cause_A}, we discussed how instruction misalignment can prevent macro-op fusion in the front-end of modern x86 processors. We now examine a distinct effect of instruction misalignment that is caused by a microcode update issued by Intel for Skylake and later processors.  We refer to this behavior as \obsB. 
Specifically, we observe that misaligning jump instructions and their predecessor can increase fetch and decode time in the front-end when the corresponding \uops{} are not present in the \uop{} cache.  
{In 2019, Intel released a microcode update to mitigate erratum SKX102, commonly referred to as the jump conditional code (JCC) erratum. This erratum can produce unpredictable behavior on certain Intel processors under specific microarchitectural conditions that involve jump instructions crossing 64-byte boundaries. In a short report, Intel states that the microcode update prevents jump instructions that cross or terminate on 32-byte boundaries from being retained in the \uop{} cache on the affected processors~\cite[Sec.~2.1~lines 1-3]{intel2024mitigationJccErratum}.} In such cases, the instruction fetch unit must retrieve the instructions from the L1I cache, fetching 16 bytes per cycle~\cite{ren2021deadUops}. If the instructions are misaligned, the fetch unit must retrieve two blocks instead of one (cf. Figure~\ref{fig:our_misalignment}). However, the macro-op queue can only fuse the instructions once both are available. 
We observe that the macro-op queue waits an additional cycle to retrieve both instructions before attempting macro-op fusion.
This introduces a timing variability of one cycle, depending on the placement of the jump instruction and its predecessor. 
Assuming that most jump instruction lengths range from 2-7 bytes and that the preceding fusible instructions are 3-5  bytes long,\footnote{We obtained these estimates from our analysis of 320 Ubuntu 24.04.2 default libraries (cf. Section~\ref{sec:performance}).} the total length of a fusible instruction pair falls within five and 12 bytes, allowing for anywhere between four and 11 alignments to cross the 32-byte boundary. Based on this distribution, and excluding the alignment causing \obsA, the likelihood of a fusible instruction pair encountering such alignment is lower-bounded by $\frac{7}{64}=10.93\%$ and upper-bounded by $\frac{21}{64}=32,81\%$. Our code in Figure~\ref{lst:instrumented_loop} mimics this distribution by selecting a four-byte preceding instruction, which aligns with the observed average, followed by a standard two-byte conditional jump.

To validate this observation, we now analyze the timing behavior of the exemplary loop in Figure~\ref{lst:instrumented_loop} 
using the same experimental setup detailed in Section~\ref{sec:root_cause_A}. In the following, we present and discuss our findings.

\vspace{0.5 em}\noindent \textbf{Detailed findings.}
We witness an increased fetch time in the front-end for \offset s 27-31, 59, and 61-63 when misaligning two macro-fusible instructions. 
Table~\ref{tab:exceprt_obs2} highlights the experimental results for selected \offset s and HPCs to validate this observation. The execution time measurements in the first row of the table indicate that \offset{} 31 exhibits an increased execution time of approximately 200 cycles compared to \offset s 0 and 26, indicating that they undergo a different treatment by the microarchitecture.

\begin{table}[tbp]
    \centering
    \caption{Events for a selected subset of \offset s and HPCs relevant for \obsB{} in the Coffee Lake microarchitecture.}
    \label{tab:exceprt_obs2}
        \scalebox{0.83}{\begin{tabular}{|p{4.0cm}|l|c|c|c|} 
             \hline
             \multirow{2}{*}{\textbf{HPC}} & \multirow{2}{*}{\textbf{event}} & \multicolumn{3}{c|}{\textbf{\offset}} \\ \cline{3-5}
              & & \textbf{0} & \textbf{26} & \textbf{31} \\
             \hline
             \texttt{rdtsc}                                 & execution cycles                & 305 & 322 & 508 \\ \hline
             \texttt{IDQ\_UOPS\_NOT\_DELIVERED.} \texttt{CYCLES\_0\_DELIV\_CORE} & IDQ stall cycles & 281 & 274 & 467 \\ \hline
             \texttt{FRONTEND\_RETIRED.}\texttt{DSB\_MISS} & \uop{} cache misses             &  74 & 230 & 230 \\ \hline
        \end{tabular}}
\end{table}

As shown in the second row of Table~\ref{tab:exceprt_obs2}, this timing discrepancy is caused by a stall in the IDQ, which serves as a \uop{} buffer between the front-end and the back-end. Specifically, the number of cycles where the IDQ does not deliver any \uops{} to the back-end remains around 280 for \offset s 0 and 26, whereas for offset 31, the HPC reports 467 stalled cycles. Note that the hardware performance counter 
\texttt{IDQ\_UOPS\_NOT\_DELIVERED.CYCLES\_0\_UOPS\_DELIV\_CORE} only counts IDQ stall cycles that are not caused by the back-end, confirming that these stalls originate in the front-end rather than in the back-end.
The third row of the table presents measurements for the \texttt{FRONTEND\_RETIRED.DSB\_MISS} HPC, which counts the number of instructions experiencing \uop{} cache misses (i.e., instructions fetched via the decode pipeline rather than the \uop{} cache). We observe an increased number of \uop{} cache misses for \offset s 26 and 31 compared to \offset{} 0. This indicates that for \offset s 26 and 31, loop instructions are fetched from the decode pipeline, whereas for \offset{} 0, they are served directly from the \uop{} cache.

Recall that { our experiments use a processor} based on the Coffee Lake microarchitecture, where the instruction fetch unit retrieves one 16-byte aligned code block per cycle. 
Figure~\ref{fig:instruction_shift} illustrates the alignment of the two instructions in the loop for \offset s 26-31. { Notably, these offsets correspond to alignments for fusible instructions that are excluded from \uop{} cache residency by Intel’s microcode update~\cite[see Fig.~1, Examples 4-7]{intel2024mitigationJccErratum} and therefore must be fetched via the decode pipeline.}
At \offset{} 26, we note that the two instructions do not cross the 16-byte aligned boundary. 
Thus, the entire loop can be fetched in a single cycle, explaining { the faster execution despite increased} \uop{} cache misses. At \offset{} 31, however, one of the instructions crosses the boundary, so they { no longer fit within} the same 16B fetch window. 

\begin{figure}
    \centering
    \includegraphics[width=0.97\linewidth]{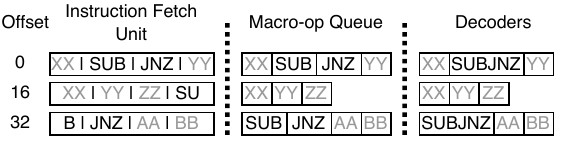}
    \caption{Overview of the decode pipeline for different placements of two fusible instructions. In the upper case, the instructions can be fetched without delay, whereas the lower case incurs a one-cycle delay for fetching the remainder of the \texttt{sub} instruction. }
    \label{fig:misalignment}
\end{figure}

Figure~\ref{fig:misalignment} illustrates how this leads to additional fetch delays in the decode pipeline. In the first scenario (top row), the \texttt{sub} and \texttt{jnz} instructions are placed within a single 16B block. The instruction fetch unit retrieves both instructions in a single cycle, delivering them together to the macro-op queue for fusion. The decoder then processes the fused instruction without additional delay.
In contrast, in the second scenario (second and third row), the \texttt{sub} instruction is split across two blocks. In this case, the instruction fetch unit cannot deliver the incomplete instruction to the macro-op queue within the same cycle. Instead, the fetch unit must wait for the subsequent block, delaying macro-op fusion by one cycle. This additional fetch cycle causes a one-cycle delay in instruction execution. This scenario occurs for nine out of 64 \offset s, i.e., with a probability of $\frac{7}{64} \leq 14.06\% \leq \frac{21}{64}$, well within the ranges obtained above.

A comprehensive view of our measurements for all 64 \offset s is presented in Figure~\ref{fig:results_all}(a-d). Here, we highlight the \offset s that prevent macro-op fusion (cf. Section~\ref{sec:root_cause_A}) in blue, while those that increase the instruction fetch delay in the front-end (cf. Section~\ref{sec:root_cause_B}) are marked in orange.

\input{exp_results_all}

\section{Impacted Architectures} 
\label{sec:reproducibility}

Our analysis in Section~\ref{sec:problem} focuses on a specific processor, namely the Intel Xeon E-2286G CPU, which implements Intel's Coffee Lake microarchitecture. We now examine the reproducibility of our findings across different processors and analyze the impact of the microarchitecture. Specifically, we performed the measurements described in Section~\ref{sec:root_cause_A} on eight different processors spanning seven distinct microarchitectures. Our experiments primarily consisted of profiling the execution time and logging HPC events for a \texttt{sub-jnz} loop with 200 iterations (cf. \Cref{lst:instrumented_loop}). The last two rows in Figure~\ref{fig:results_all} present an overview of the timing measurements, while Table~\ref{tab:uarchs} summarizes the HPC data for selected offsets discussed in our analysis in Section~\ref{sec:problem}. 
Notice that variations in the naming of HPCs stem from slight naming variances on different architectures.  
Based on our findings, we classify the observed behavior into three distinct categories of reproducibility, detailed in the following sections.

\subsection{Skylake, Coffee Lake, and Kaby Lake (R)}
\label{sec:sky_coffee_kaby}

We start by examining the processors that exhibit timing behavior consistent with the Intel Xeon E-2286G CPU used in our primary experiments. Specifically, Figures~\ref{fig:coffee_rub} to \ref{fig:sky_rub} show that the Intel Xeon E-2286G, Xeon E-2176G, Core i7-8550U, Core i5-8350U, and Core i7-6300U CPUs all show similar timing variations. To validate that these variations come from \obsA{} and \obsB{} effects, we analyzed the HPC data, as summarized in Table~\ref{tab:uarchs}. 

All examined CPUs show an increase of approximately 200 \uops{} at \offset{} 60 due to the lack of macro-op fusion for this \offset{}  (cf. Section~\ref{sec:root_cause_A}). Moreover, at \offset s 26 and 31, we observe an increase of approximately 180 \uop{} cache misses, while only \offset{} 31 results in IDQ stalls caused by increased instruction fetch latency (cf. Section~\ref{sec:root_cause_B}). Measured differences in execution cycles across these processors are primarily due to variations in their operating frequencies, as each CPU was tested at its respective base frequency.

These processors implement one of three closely related microarchitectures: Skylake, Coffee Lake, and Kaby Lake. Notably, Kaby Lake and Coffee Lake are distinct successors of Skylake but do not introduce significant front-end modifications. All three microarchitectures are affected by the Intel microcode update that modifies the \uop{} cache behavior~\cite{intel2024mitigationJccErratum}. Moreover, this microcode update affects several additional CPUs based on Amber Lake, Cascade Lake, Comet Lake, and Whiskey Lake. While we do not have access to these specific processors for testing, we expect that these microarchitectures are equally subject to \obsA{} and \obsB.

Note that the LSD, which caches small loops directly within the IDQ, was disabled by a microcode update on the Skylake, Coffee Lake, and Kaby Lake processors we analyzed, due to erratum SKL150~\cite{erratumSKL150}. However, the LSD remains active on Amber Lake, Cascade Lake, Comet Lake, and Whiskey Lake processors. Since the LSD bypasses both the \uop{} cache (relevant to \obsA) and the decode pipeline (relevant to \obsB), its use can mitigate the timing effects we observe. Nonetheless, we expect the LSD to only partially reduce the impact, as it does not cover all loops.

\subsection{Ice Lake}

We now analyze another Intel microarchitecture, namely Ice Lake. This architecture is particularly interesting because it is not listed among the affected microarchitectures in~\cite{intel2024mitigationJccErratum}. At first glance, the timing behavior of the profiled Xeon Gold 5315Y CPU appears similar to the patterns observed in Section~\ref{sec:sky_coffee_kaby}. Specifically, we measure an additional 200 cycles in the execution time for \offset s 59 to 63. However, a detailed examination of the corresponding HPC measurements in Table~\ref{tab:uarchs} reveals partially deviating root causes. 

We find that approximately 140 additional \uops{} are issued at \offset{} 60, aligning with our observations from Section~\ref{sec:root_cause_A} and confirming that \obsA{} remains effective on this architecture. When considering \obsB{}, although we witness an additional 200 IDQ stalls at \offset s 59 and 61-63, our measurements suggest that these do not result from an increased number of \uop{} cache misses.

Note that processors based on the Ice Lake microarchitecture were introduced in 2021---two years after the publication of~\cite{intel2024mitigationJccErratum}. We hypothesize that Ice Lake (and subsequent microarchitectures) addresses the abnormality arising from the microcode update at a hardware level, such that the alignment previously associated with \obsB{} no longer prevents the corresponding \uops{} from being cached in the \uop{} cache; however, the IDQ stalls remain as an artifact.
The example of Ice Lake highlights that the observed timing variabilities are highly sensitive to the microarchitectural specifics of the front-end, suggesting that other front-end changes, such as a modified fetch window size in the Lion Cove micro-architecture, also influence how \obsA{} and \obsB{} manifest. Consequently, individual inspection of each architecture is required to verify the manifestation of these issues.

\newcommand{\rdtscEntry}[5]{ & & & \texttt{rdtsc} & #1 & #2 & #3 & #4 & #5 }
\newcommand{\issuedEntry}[5]{ & & & \texttt{UOPS\_ISSUED.ALL} & #1 & #2 & #3 & #4 & #5 }
\newcommand{\dsbEntry}[5]{ & & & \texttt{FRONTEND\_RETIRED.DSB\_MISS} & #1 & #2 & #3 & #4 & #5 }
\newcommand{\idqEntry}[5]{ & & & \texttt{IDQ\_UOPS\_NOT\_DELIVERED. CYCLE\_0\_UOPS\_DELIV\_CORE} & #1 & #2 & #3 & #4 & #5 }

\newcommand{\issuedEntryIce}[5]{ & & & \texttt{UOPS\_ISSUED.ANY} & #1 & #2 & #3 & #4 & #5 }

\newcommand{\issuedEntryAMD}[5]{ & & & \texttt{RETIRED\_FUSED\_INSTRUCTIONS} & #1 & #2 & #3 & #4 & #5 }
\newcommand{\dsbEntryAMD}[5]{ & & & \texttt{OP\_CACHE\_HIT\_MISS.OC\_HIT} & #1 & #2 & #3 & #4 & #5 }
\newcommand{\dsbEntryAMDs}[5]{ & & & \texttt{UOPS\_DISPATCHED\_FROM\_DECODER. OPCACHE\_DISPATCHED} & #1 & #2 & #3 & #4 & #5 }
\newcommand{\idqEntryAMD}[5]{ & & & \texttt{UOPS\_QUEUE\_EMPTY} & #1 & #2 & #3 & #4 & #5 }

\begin{table*}[tbp]
    \centering
    \caption{Number of events recorded for a selected subset of \offset s and HPCs relevant to \obsA{} and \obsB{} across various Intel and AMD processors. Values that are significantly higher at a specific \offset{} compared to other \offset s are highlighted in \textbf{bold}.}
    \label{tab:uarchs}
    \footnotesize
    \scalebox{0.97}{
        \begin{tabular}{cccp{3.2cm}|c|cc|cc p{0.05cm} cccp{3.6cm}|c|cc|cc}
            
            \cmidrule[0.75pt]{1-9}\cmidrule[0.75pt]{11-19}
             
            & & & Finding &    & \multicolumn{2}{c|}{\obsA} & \multicolumn{2}{c}{\obsB}                 & &     & & & Finding &    & \multicolumn{2}{c|}{\obsA} & \multicolumn{2}{c}{\obsB} \\
            & & & Offset  &  0 & 32 & 60                    & 26 & 31                                   & &     & & & Offset  &  0 & 32 & 60                    & 26 & 31 \\
             
            \cmidrule{1-9}\cmidrule{11-19}

            \parbox[t]{1mm}{\multirow{24}{*}{\rotatebox[origin=c]{90}{\shortstack{Intel}}}} &
            \parbox[t]{1mm}{\multirow{12}{*}{\rotatebox[origin=c]{90}{\shortstack{Coffee Lake}}}} &
            \parbox[t]{1mm}{\multirow{6}{*}{\rotatebox[origin=c]{90}{\shortstack{Xeon E-2286G}}}}
                & \multicolumn{6}{c}{}                                                                  & &     \parbox[t]{1mm}{\multirow{12}{*}{\rotatebox[origin=c]{90}{\shortstack{Intel}}}} &
                                                                                                                \parbox[t]{1mm}{\multirow{6}{*}{\rotatebox[origin=c]{90}{\shortstack{Skylake}}}} &
                                                                                                                \parbox[t]{1mm}{\multirow{6}{*}{\rotatebox[origin=c]{90}{\shortstack{i5-6300U}}}}
                                                                                                                    & \multicolumn{6}{c}{} \\
            \rdtscEntry{305}{307}{\textbf{495}}{322}{\textbf{508}}                                      & &     \rdtscEntry{272}{271}{\textbf{434}}{266}{\textbf{431}} \\
            \issuedEntry{611}{616}{\textbf{780}}{615}{612}                                              & &     \issuedEntry{754}{753}{\textbf{936}}{754}{750} \\
            \dsbEntry{74}{74}{65}{\textbf{230}}{\textbf{230}}                                           & &     \dsbEntry{53}{54}{52}{\textbf{230}}{\textbf{231}} \\
            \idqEntry{281}{286}{300}{274}{\textbf{467}}                                                 & &     \idqEntry{322}{316}{340}{314}{\textbf{503}} \\
            
            \cmidrule{3-9}                                                                                      \cmidrule{12-19}

             &
             &
            \parbox[t]{1mm}{\multirow{6}{*}{\rotatebox[origin=c]{90}{\shortstack{Xeon E-2176G}}}}
                & \multicolumn{6}{c}{}                                                                  & &      &
                                                                                                                \parbox[t]{1mm}{\multirow{6}{*}{\rotatebox[origin=c]{90}{\shortstack{Ice Lake}}}} &
                                                                                                                \parbox[t]{1mm}{\multirow{6}{*}{\rotatebox[origin=c]{90}{\shortstack{Xeon Gold 5315Y}}}}
                                                                                                                    & \multicolumn{4}{c}{} & 58 & 63 \\
            \rdtscEntry{261}{257}{\textbf{417}}{259}{\textbf{410}}                                      & &     \rdtscEntry{274}{275}{\textbf{451}}{307}{\textbf{509}} \\
            \issuedEntry{684}{682}{\textbf{872}}{684}{683}                                              & &     \issuedEntryIce{600}{611}{\textbf{742}}{606}{601} \\
            \dsbEntry{47}{50}{46}{\textbf{229}}{\textbf{230}}                                           & &     \dsbEntry{35}{34}{70}{39}{51} \\
            \idqEntry{335}{269}{260}{257}{\textbf{439}}                                                 & &     \idqEntry{182}{183}{210}{199}{\textbf{397}} \\
            & & & \multicolumn{6}{c}{}                                                                  & &     & & & \multicolumn{6}{c}{} \\
            
            \cmidrule{2-9}                                                                                      \cmidrule{11-19}

             &
            \parbox[t]{1mm}{\multirow{6}{*}{\rotatebox[origin=c]{90}{\shortstack{Kaby Lake}}}} &
            \parbox[t]{1mm}{\multirow{6}{*}{\rotatebox[origin=c]{90}{\shortstack{i7-8550U}}}}
                & \multicolumn{6}{c}{}                                                                  & &     \parbox[t]{1mm}{\multirow{12}{*}{\rotatebox[origin=c]{90}{\shortstack{AMD}}}} &
                                                                                                                \parbox[t]{1mm}{\multirow{6}{*}{\rotatebox[origin=c]{90}{\shortstack{Zen 2}}}} &
                                                                                                                \parbox[t]{1mm}{\multirow{6}{*}{\rotatebox[origin=c]{90}{\shortstack{EPYC 7262}}}}
                                                                                                                    & \multicolumn{4}{c}{} & 58 & 63 \\
            \rdtscEntry{181}{177}{\textbf{289}}{180}{\textbf{284}}                                      & &     \rdtscEntry{325}{336}{\textbf{508}}{332}{\textbf{512}} \\
            \issuedEntry{730}{738}{\textbf{919}}{728}{728}                                              & &     \issuedEntryAMD{17}{17}{17}{16}{17} \\
            \dsbEntry{50}{48}{49}{\textbf{231}}{\textbf{232}}                                           & &     \dsbEntryAMDs{783}{790}{756}{746}{760} \\
            \idqEntry{290}{321}{314}{383}{\textbf{606}}                                                 & &     \idqEntryAMD{182}{182}{176}{203}{\textbf{381}} \\
            
            \cmidrule{2-9}                                                                                      \cmidrule{12-19}

             &
            \parbox[t]{1mm}{\multirow{6}{*}{\rotatebox[origin=c]{90}{\shortstack{Kaby Lake R}}}} &
            \parbox[t]{1mm}{\multirow{6}{*}{\rotatebox[origin=c]{90}{\shortstack{i5-8350U}}}}
                & \multicolumn{6}{c}{}                                                                  & &      &
                                                                                                                \parbox[t]{1mm}{\multirow{6}{*}{\rotatebox[origin=c]{90}{\shortstack{Zen 3}}}} &
                                                                                                                \parbox[t]{1mm}{\multirow{6}{*}{\rotatebox[origin=c]{90}{\shortstack{EPYC 7313P}}}}
                                                                                                                    & \multicolumn{4}{c}{} & 58 & 63 \\
            \rdtscEntry{179}{178}{\textbf{287}}{180}{\textbf{279}}                                      & &     \rdtscEntry{269}{281}{\textbf{449}}{269}{\textbf{449}} \\
            \issuedEntry{700}{700}{\textbf{882}}{703}{698}                                              & &     \issuedEntryAMD{227}{227}{226}{227}{227} \\
            \dsbEntry{43}{46}{46}{\textbf{228}}{\textbf{228}}                                           & &     \dsbEntryAMD{254}{241}{\textbf{451}}{257}{\textbf{452}} \\
            \idqEntry{292}{287}{316}{300}{\textbf{496}}                                                 & &     \idqEntryAMD{196}{225}{200}{191}{202} \\
            
            \cmidrule[0.75pt]{1-9}                                                                              \cmidrule[0.75pt]{11-19}
        
    \end{tabular}}
\end{table*}

\subsection{The Case of AMD}
\label{sec:amd}

We now present a set of experiments to verify the occurrence of \obsA{} on AMD's Zen 2 and Zen 3 microarchitectures. We note that \obsB{}, which is a result of an Intel-specific microupdate~\cite{intel2024mitigationJccErratum}, does not apply to AMD. 

As shown in Figures~\ref{fig:zen2_nec} and \ref{fig:zen3_rub}, the profiled EPYC 7262 and EPYC 7313P processors exhibit execution time patterns similar to those observed in Intel microarchitectures. Specifically, we observe an additional 180-200 cycle delay for \offset s 59 to 63. 
Note that, in Zen 2, (i) this behavior is independent of the number of fused instructions (the \texttt{RETIRED\_FUSED\_INSTRUCTIONS} HPC consistently reports only 17 fused instructions across all \offset s, aligning with the fact that Zen 2 exclusively fuses conditional jumps with \texttt{test} and \texttt{cmp} instructions~\cite{agner2024microarchitecture}.), and (ii) the number of \uops{} dispatched from the \uop{} cache remains constant, confirming that all \uops{} are retrieved from the \uop{} cache. However, loops with $\leq$5 instructions confined to 64B blocks can take a fast path, executing in one cycle per iteration (offsets 0-58), whereas crossing the boundary (offsets 59-63) adds a one-cycle penalty, since only one taken branch is executed every two cycles~\cite{agner2024microarchitecture}. This effect mimics the behavior of \obsA{} in x86 architectures.

In contrast, Zen 3 expands the set of fusible instructions such that we observe macro-op fusion for all \offset s. That is, the \texttt{sub} and \texttt{jnz} are macro-fused and translated into a single \uop{}, occupying only one slot in a single \uop{} cache set. However, the \linebreak \texttt{OP\_CACHE\_HIT\_MISS.OC\_HIT} HPC records approximately 200 additional hits in the \uop{} cache for \offset s 60 and 63, suggesting that Zen 3 still accesses two distinct \uop{} cache sets. This introduces an additional cycle delay, mimicking the impact of \obsA.

\section{Abnormal Conditional Jumps in the Wild}
\label{sec:performance}

To understand the real-world implications of our findings, we assess the prevalence of \obsA{} and \obsB{} in 666 real-world binaries, ranging from popular cryptographic libraries to those shipped with current operating systems.

\subsection{Methodology}

To cover a wide range of real-world applications, we analyzed the binaries of libraries shipped with recent versions of two widely used operating systems: Ubuntu 24.04.2 and Windows 10 Pro 22H2. More specifically, we analyzed all the shared libraries installed per default on a fresh Ubuntu 24.04.2 installation. Moreover, we installed and analyzed the \texttt{apt} package \texttt{build-essential}, which includes a set of basic development tools, such as \texttt{gcc} and \texttt{make}. We retrieved the binaries of these libraries by executing the \texttt{/sbin/ldconfig -p} command, yielding 320 results. Similarly, we analyzed all the shared libraries installed per default on a fresh Windows 10 Pro 22H2 installation. Here, we retrieved the binaries of these libraries using a \texttt{python} script that extracts all DLL files in the ``\texttt{C:\textbackslash Program Files (x86)}'' folder, yielding 323 results.

Besides analyzing the 320 libraries in Ubuntu 24.04.2 and the 323 libraries in Windows 10 Pro 22H2, we also analyzed a set of {23} cryptographic libraries {(resulting in a total of 666 libraries)}, which are typically highly optimized for performance and security. Our selected cryptographic libraries are taken from~\cite{jancar2024theyreNotThatHardToMitigate}---which is, at the time of writing, one of the most extensive analyses of constant time behavior in cryptographic libraries. We focused our analysis on open-source libraries for the x86 microarchitecture and excluded libraries for other architectures, such as Arm, from our analysis. Furthermore, we did not analyze libraries written in languages other than C/C++, { resulting in} 23 cryptographic libraries.

\begin{table}[]
	\centering
	\small
	\begin{tabular}{|l|c|}
		\hline
		\textbf{First instruction}  & \textbf{Fusible conditional jumps} \\ \hline
		cmp   & jz, jc, jb, ja, jl, jg, je \\\hline
		add   & jz, jc, jb, ja, jl, jg, je \\\hline
		sub   & jz, jc, jb, ja, jl, jg, je \\\hline
		inc   & jz, jl, jg, je \\\hline
		dec   & jz, jl, jg, je \\\hline
		test  & jz, jc, jb, ja, jl, jg, js, jp, jo, je \\\hline
		and   & jz, jc, jb, ja, jl, jg, js, jp, jo, je \\\hline
	\end{tabular}
	\vspace{0.2cm}
	\caption{Fusible instruction pairs. The negated jumps, e.g., \texttt{jnz} for \texttt{jz}, are equally macro-fusible with the respective first instruction.} 
	\label{tab:fusible_jumps_list}
\end{table}

\subsection{Static Analysis}

Our evaluation is based on a static analysis of the libraries' binaries.  
We disassembled the binaries and extracted all conditional jumps together with their preceding instructions, recording each instruction's address and length.
Next, we analyzed if the extracted instruction pairs can be macro-fused. At this stage, we did not yet consider the instruction placement, but only whether the two instructions are compatible. Therefore, we used the table of macro-fusible instructions from~\cite[Sec. 9.6]{agner2024microarchitecture} and extended it with an additional conditional jump: we found that the \texttt{je} conditional jump instruction can also be fused with preceding arithmetic, logic, and compare instructions. Table~\ref{tab:fusible_jumps_list} provides an overview of fusible instruction pairs. Additionally, macro-op fusion is constrained by the type of operands of the first instruction. In Skylake-based microarchitectures, an instruction cannot be fused with a subsequent conditional jump: (i) if one of the operands uses \texttt{rip}-relative addressing, (ii) if the destination operand is a memory operand, or (iii) if the instruction has both an immediate and a memory source operand. Our static analysis accounts for this by removing instructions matching these criteria from the set of fusible instruction pairs.

Recall that macro-op fusion fails if the op and the jump of such a pair are in distinct L1I cache sets, i.e., the jump instruction starts at offset 64 (cf. Section~\ref{sec:root_cause_A}). To analyze how many of the jumps fall into \obsA{}, we analyzed all fusible operation-jump pairs and counted for how many of those the address of the conditional jump is 64B aligned.

On the other hand, macro-fusible instruction pairs are not cached and incur a delay in the instruction fetch unit if they span a 32B boundary but are not subject to \obsA{} (cf. Section~\ref{sec:root_cause_B}). Therefore, we analyzed whether the starting address of the first instruction was in a different 32B-aligned block than the last byte of the conditional jump.

Note that our analysis adopted a conservative approach by considering only the conditional jumps listed in Table~\ref{tab:fusible_jumps_list} as candidates for macro-op fusion. While this selection aligns with prior work~\cite{agner2024microarchitecture,wang2023crossCoreMacroOpFusion}, it is not exhaustive as it does not consider additional conditional jump instructions, such as \texttt{jae} or \texttt{jnge}. Given that these additional instructions are classified as conditional jumps, this effectively reduces the fraction of \obsA{} and \obsB{} reported in our analysis. Similarly, macro-op fusion does not allow for \texttt{rip}-relative memory accesses in the first instruction of a fusible instruction pair. While the \texttt{rip} register is permitted as an operand, our analysis excluded all instructions that access the \texttt{rip}. Due to the conservative nature of our approach, we anticipate a low rate of false positives.

\subsection{Evaluation Results}
\label{sec:perf_eval_results}

Due to space limitations, we could not include the detailed results of all 666 libraries in this paper. Instead, Table~\ref{tab:performance_analysis} presents a representative summary of our findings.
The complete list of libraries and our full evaluation results can be accessed at~\cite{our_artifact}. 
We summarize our findings below.

\begin{itemize}
    \item Out of the 320 Ubuntu 24.04.2 libraries that we investigated, we identified 4,100,969 conditional jumps.
    From these jumps, 1.28\% were subject to \obsA{}, and 14.95\% were caused by \obsB. 
    Overall, the average proportion of \obsA{} occurrences falls slightly below the expected value. 
    In contrast, the proportion of \obsB{} occurrences remains above the lower bound of 10.93\% (see Section~\ref{sec:root_cause_B}) for 309 out of the 320 libraries, although our analysis uses a conservative heuristic.
    Notably, only a single library in our data set exhibited neither \obsA{} nor \obsB{} jumps.
    
    \item In our analysis of 323 Windows 10 Pro 22H2 libraries, we identified a total of 18,769,762 conditional jumps, with 1.15\% resulting from \obsA{} and 13.69\% from \obsB.
    Among the analyzed libraries, 42 libraries do not contain any \obsA{} or \obsB{} jumps. Of these, 27 libraries contained no conditional jumps at all, while the remaining 15 may have leveraged the jump alignment optimization discussed in Section~\ref{sec:mitigation}.
    
    \item Among the 23 cryptographic libraries we analyzed, we identified 396,093 conditional jumps. Among these jumps, 1.17\% were caused by \obsA, while 15.08\% were attributed to \obsB. Consistent with the libraries in Ubuntu 24.04.2 and Windows 10 Pro 22H2, these results fall slightly below the expected value for \obsA{}, and confirm our lower bound of 10.93\% for \obsB. 
  \end{itemize}

\noindent \textbf{Summary of results: }
Our empirical results validate our analysis in Section~\ref{sec:problem}. Specifically, they confirm that conditional jumps are likely to exhibit abnormal execution timing in more than 15.1\% (averaged over all occurrences of \obsA{} and \obsB{} across all 666 libraries that we investigated). Notably, while the expected proportion of \obsA{} occurrences is 1.56\% (cf. Section~\ref{sec:root_cause_A}), our empirical measurements show that only 1.17\% of conditional jumps are affected by \obsA. This discrepancy arises because our analysis in Section~\ref{sec:root_cause_A} conservatively assumes that all conditional jumps are macro-fusible. On the other hand, the measured proportion for \obsB{} occurrences is 13.93\%, which aligns well with the expected range of 10.93 to 32.81\% outlined in Section~\ref{sec:root_cause_B}. In particular, this value closely matches the 14.06\% measured in our sample in Section~\ref{sec:root_cause_B}.

\definecolor{gainsboro}{rgb}{0.86, 0.86, 0.86}

\newcommand{\cryptoLibEntry}[5]{ #1 & #2 & #3\% & #4\% }
\newcommand{\OSEntry}[4]{ #1 & #2 & #3\% & #4\% }

\begin{table*}[tbp]
	\centering
    \caption{Excerpt from our static analysis of 666 libraries across Ubuntu 24.04.2, Windows 10 Pro 22H2, and cryptographic libraries. For each library, we report the total number of conditional jumps and the proportion of jumps affected by \obsA{} and \obsB{}.
    }
    \label{tab:performance_analysis}
    \small
    \scalebox{1.0}{
    \begin{tabular}{|l|P{1.6cm}|c|c?l|P{1.6cm}|c|c|}

    \hline
    \textbf{Library} & \textbf{Conditional jumps} & \textbf{\obsA} & \textbf{\obsB} & \textbf{Library} & \textbf{Conditional jumps} & \textbf{\obsA} & \textbf{\obsB} \\
    \hline

    \rowcolor{gainsboro}
    \multicolumn{8}{|c|}{\textbf{Ubuntu 24.04.2}} \\ \hline
    \OSEntry{libusb-1.0.so.0 }{1474}{1.02}{16.76}               & \OSEntry{libkrb5.so.3 }{7564}{1.20}{15.04} \\ \hline
    \OSEntry{liburcu-mb.so.8 }{363}{1.65}{12.95}                & \OSEntry{libjbig.so.0 }{710}{0.28}{11.13} \\ \hline 
    \OSEntry{libthread\_db.so }{448}{1.34}{15.18}               & \OSEntry{libidn2.so.0 }{447}{1.79}{20.36} \\ \hline 
    \OSEntry{libseccomp.so.2 }{1252}{1.60}{18.13}                & \OSEntry{libhogweed.so.6 }{891}{1.12}{14.48} \\ \hline 
    \OSEntry{libquadmath.so.0 }{3553}{1.07}{17.62}              & \OSEntry{libguestStoreClient.so.0 }{1884}{0.96}{12.05} \\ \hline
    \OSEntry{libply-boot-client.so.5 }{135}{3.70}{13.33}         & \OSEntry{libfreetype.so.6 }{15737}{1.05}{15.85} \\ \hline
    \OSEntry{libpam.so.0 }{850}{0.94}{16.47}                    & \OSEntry{libestr.so.0 }{86}{1.16}{9.30} \\ \hline 
    \OSEntry{libncurses.so.6 }{2868}{1.43}{15.17}               & \OSEntry{libdb-5.3.so }{35937}{1.17}{18.34} \\ \hline 
    \OSEntry{libmpc.so.3 }{1425}{1.12}{20.63}                   & \OSEntry{libcap-ng.so.0 }{305}{1.64}{18.36} \\ \hline 
    \OSEntry{libm.so.6 }{7548}{0.79}{10.73}                     & \OSEntry{libc\_malloc\_debug.so }{726}{0.83}{13.36} \\ \hline

    \rowcolor{gainsboro}
    \multicolumn{8}{|c|}{\textbf{Windows 10 Pro 22H2}} \\ \hline
    \OSEntry{wdag.dll}{3316}{1.09}{11.73}                   & \OSEntry{vccorlib140.dll}{4702}{1.62}{16.55} \\ \hline
    \OSEntry{ink-rtscom.dll}{2205}{1.09}{11.75}             & \OSEntry{wns\_push\_client.dll}{6929}{0.92}{12.11} \\ \hline
    \OSEntry{msdadc.dll}{92}{1.09}{8.70}                     & \OSEntry{msedgeupdateres\_da.dll}{34}{2.94}{0.00} \\ \hline
    \OSEntry{msdatl3.dll}{3453}{0.55}{5.50}                  & \OSEntry{msedgeupdateres\_mt.dll}{33}{3.03}{0.00} \\ \hline
    \OSEntry{hmmapi.dll}{353}{2.27}{8.22}                   & \OSEntry{libEGL.dll}{8765}{1.15}{13.83} \\ \hline
    \OSEntry{libEGL.dll}{8175}{1.57}{13.74}                 & \OSEntry{msedge\_elf.dll}{55243}{1.17}{13.95} \\ \hline
    \OSEntry{oneds.dll}{61437}{1.25}{15.29}                 & \OSEntry{msvcp140.dll}{8409}{1.21}{12.74} \\ \hline
    \OSEntry{wns\_push\_client.dll}{11618}{1.29}{13.61}     & \OSEntry{wdag.dll}{3316}{1.09}{11.73} \\ \hline
    \OSEntry{mip\_core.dll}{44430}{1.36}{14.50}              & \OSEntry{ie\_to\_edge\_bho\_64.dll}{11162}{1.33}{13.91} \\ \hline
    \OSEntry{concrt140.dll}{4024}{1.24}{13.34}              & \OSEntry{EppManifest.dll}{0}{0.00}{0.00} \\ \hline

    \rowcolor{gainsboro}
    \multicolumn{8}{|c|}{\textbf{Cryptographic libraries}} \\ \hline
    \cryptoLibEntry{libgcrypt 1.11.0}{9568}{1.09}{16.16}{O2}            & \cryptoLibEntry{tiny-AES-c 1.0.0}{20}{0.00}{10.00}{Os} \\ \hline
    \cryptoLibEntry{OpenSSL 3.4.0}{63794}{1.20}{16.71}{O3}              & \cryptoLibEntry{NSS 3.107}{10564}{1.12}{15.23}{O2} \\ \hline
    \cryptoLibEntry{LibreSSL 4.0.0}{16730}{1.16}{15.36}{O2}             & \cryptoLibEntry{libtomcrypt 1.18.2}{7152}{1.06}{15.86}{O3} \\ \hline
    \cryptoLibEntry{BoringSSL 0.20250114.0}{31177}{1.26}{16.00}{O2/O3}  & \cryptoLibEntry{Nettle 3.10.1}{2277}{1.14}{13.92}{O2} \\ \hline
    \cryptoLibEntry{BearSSL 0.6}{1636}{1.47}{12.84}{Os}                 & \cryptoLibEntry{MS SymCrypt 103.6.0}{4681}{2.16}{15.30}{O3} \\ \hline
    \cryptoLibEntry{Botan 3.6.1}{72328}{1.25}{16.07}{O3}                & \cryptoLibEntry{Intel IPP crypto 1.0.1}{83515}{1.01}{11.91}{O3/32} \\ \hline
    \cryptoLibEntry{Crypto++ 8.9}{33829}{1.07}{15.80}{O3}               & \cryptoLibEntry{NaCl 20110221}{697}{0.57}{12.20}{O3} \\ \hline
    \cryptoLibEntry{wolfSSL 5.7.6}{10663}{1.32}{15.82}{O2}              & \cryptoLibEntry{libsecp256k1 0.6.0}{695}{0.72}{16.12}{O2} \\ \hline
    \cryptoLibEntry{mbedTLS 3.6.2}{10157}{1.19}{16.04}{O2}              & \cryptoLibEntry{libsodium 1.0.20}{1871}{1.50}{12.93}{O3} \\ \hline
    \cryptoLibEntry{Amazon s2n 1.5.11}{9044}{1.36}{17.80}{O2}           & \cryptoLibEntry{SPHINCS+}{903}{1.33}{17.17}{O3} \\ \hline 
    \cryptoLibEntry{GnuTLS 3.7.11}{23587}{1.17}{14.27}{O2}              & \cryptoLibEntry{Monocypher 4.0.2}{264}{1.89}{11.74}{O3} \\ \hline 
    \cryptoLibEntry{PQCrypto-SIDH 3.5.1}{941}{1.38}{14.56}{O3}          & & & & \\ \hline
    
    \hline
    
	\end{tabular}
    }
\end{table*}

\section{Mitigation \& Significance}
\label{sec:mitigation_significance}

In this section, we explore strategies to mitigate the alignment conditions responsible for \obsA{} and \obsB{}, and quantify their impact across various real-world libraries.

\subsection{Aligning Macro-Fusible Instructions}
\label{sec:mitigation}

We now discuss strategies to mitigate the effects of \obsA{} and \obsB{} on the affected microarchitectures. As discussed in Section~\ref{sec:root_cause_B}, a prerequisite for \obsB{} is that the fusible instruction pair is not cached in the \uop{} cache, a consequence of the microcode update~\cite{intel2024mitigationJccErratum}. To compensate for the performance impact of this mitigation, one can rely on a software-based solution (as proposed by Intel in~\cite{intel2024mitigationJccErratum}). For example, the GNU Assembler introduced three new optimization flags, designed to align jump instructions to a 32-byte boundary by inserting segment prefixes or \texttt{NOP} instructions without altering code semantics. Among these flags, \texttt{-malign-branch-boundary} defines the alignment boundary, \texttt{-malign-branch} specifies which instructions affected by the microcode update mitigation to align, and \texttt{-malign-branch-prefix} sets the maximum number of prefix bytes allowed for alignment. Additionally, the \texttt{-mbranches-within-32B-boundaries} configures default values for these three options, such that all conditional, unconditional, and fused jump instructions do not span a 32B boundary if this can be prevented with a maximum of five prefix bytes.

Note that this mitigation strategy successfully prevents \emph{all} fusible instruction pairs from crossing the 32B boundary, thereby mitigating both \obsA{} and \obsB. Given that this strategy prevents uncached macro-fusible instruction pairs, it naturally prevents \obsB{} but also mitigates \obsA{} as a by-product. Specifically, \obsA{} requires conditional jumps to be 64B-aligned. Preventing instruction pairs from spanning a 32B boundary indirectly disrupts this alignment, eliminating the conditions necessary for \obsA.

\subsection{Poor Visibility \& Performance Impact}
\label{sec:browsers}

Although Intel has proposed a mitigation for \obsB{}, our analysis (cf. Table~\ref{tab:performance_analysis}) reveals that \emph{more than 600 of the libraries that we investigated in Windows 10 Pro 22H2 or Ubuntu 24.04.2 do not employ it to deter \obsA{} and \obsB}. 
This confirms that Intel's mitigation strategy is not easily accessible to practitioners. 
Moreover, { documentation is sparse on this topic: GCC's \texttt{-Wa,-mbranches-\allowbreak{}within-32B-boundaries} flag}
is neither listed in GCC's option summary~\cite{gccOptionSummary} nor mentioned in the GNU Assembler manual page~\cite{asManPage}.

To assess the effectiveness of this mitigation, we quantified its impact on both performance and code size across a range of cryptographic libraries. These libraries were selected because their source code and build configurations are publicly available, allowing controlled experimentation. We included only those libraries from \Cref{sec:perf_eval_results} that offer built-in performance benchmarks, ensuring we measure metrics deemed relevant by their maintainers. Each benchmark was run five times with and without the mitigation, and we report average results. All experiments were executed on an isolated core of an Intel Xeon E-2286G CPU, which is based on the Coffee Lake microarchitecture, with 128GB RAM running Ubuntu 20.04 LTS.  We then used the \texttt{size} command-line tool to assess changes in binary size due to the added padding.

The \texttt{-Wa,-mbranches-within-32B-boundaries} compiler flag implicitly sets default values for the \texttt{-malign-branch-boundary}, \texttt{-malign-branch}, and \texttt{-malign-branch-prefix} flags. Specifically:

\begin{itemize}
   \item \texttt{-malign-branch-boundary} aligns branches to a 32-byte boundary, which is sufficient to mitigate both \obsA{} and \obsB{} without the excessive padding that a 64-byte boundary would entail.
    \item \texttt{-malign-branch} determines which types of branches are aligned. By default, the compiler aligns conditional, fused, and unconditional jumps. We found that aligning only fused jumps---i.e., the jump type affected by the \obsA{}- and \obsB-induced timing variability---is not supported by the compiler. Therefore, we experimented with configurations that either include or exclude unconditional jumps.
    \item \texttt{-malign-branch-prefix} sets the maximum number of prefix bytes (NOPs) that are used for alignment, limited to 5. We varied this value between 1 and 5 bytes, as a value of 0 effectively disables the mitigation.
\end{itemize}

\begin{table}[tbp]
\centering
\caption{Performance speedup for cryptographic libraries with the build-time mitigation. The code size overhead is reported in brackets for each library. Results are shown across various aligned jump types and prefix sizes, and reported relative (\%) to the default build of the library. Negative values indicate slowdowns. The left column indicates results for the default configuration of the mitigation as suggested by Intel.}  
\label{tab:performance_impact}
\scalebox{0.89}{
	\begin{tabular}{|l|cHHH|cHH|cHHHHH|}
		\hline
		\textbf{Branch} &\multicolumn{7}{c|}{\textbf{\texttt{jcc+fused+jmp}}} & \multicolumn{5}{c}{\textbf{\texttt{jcc+fused}}} &  \\ \cline{1-13}
		
		\textbf{Prefix} &  \multicolumn{3}{c}{\textbf{5}} & \textbf{4} & \textbf{3} & \textbf{2} & \textbf{1} & \textbf{5} & \textbf{4} & \textbf{3} & \textbf{2} & \textbf{1} & \\ \hline
		
		BearSSL & 9.62 (1.72) & 9.69 (1.72) & 9.49 (1.72) & 9.03 (1.72) & 9.73 (1.72) & 9.74 (1.72) & 9.29 (1.72) & 10.29 (1.60) & 10.06 (1.60) & 10.17 (1.60) & {10.54} (1.60) & 10.15 (1.60) & 10.54 (1.60) \\ \hline
		
		BoringSSL & 0.83 (1.62) & 0.93 (2.31) & 1.16 (3.41) & 0.79 (1.62) & 0.78 (1.62) & 0.95 (1.62) & {1.23} (1.62) & 0.34 (1.37) & 0.51 (1.37) & 0.26 (1.37) & 0.29 (1.37) & 0.58 (1.37) & 1.23 (1.62) \\ \hline
		
		Botan & 1.31 (1.55) & 1.27 (2.19) & -0.93 (3.37) & 0.18 (1.55) & 0.31 (1.55) & -1.34 (1.55) & -0.68 (1.55) & {1.74} (1.28) & -0.06 (1.28) & 0.16 (1.28) & 0.61 (1.28) & -0.97 (1.28) & 1.74 (1.28) \\ \hline
		
		Crypto++ & 0.15 (0.83) & 0.16 (1.19) & 0.16 (1.76) & 0.11 (0.83) & {0.18} (0.83) & 0.11 (0.83) & 0.1 (0.83) & 0.08 (0.7) & 0.06 (0.7) & 0.02 (0.7) & -0.05 (0.7) & 0.07 (0.7) & 0.18 (0.83) \\ \hline
		
		GnuTLS  & 0.25 (0.91) & 0.14 (0.91) & 1.85 (0.91) & {3.75} (0.91) & 2.97 (0.91) & 2.2 (0.91) & 0.98 (0.9) & 1.84 (0.71) & -0.09 (0.71) & -1.12 (0.71) & -0.96 (0.71) & -0.32 (0.71) & 3.75 (0.91) \\ \hline
		
		libgcrypt & 0.04 (0.81) & {1.51} (1.42) & -0.01 (2.59) & 0.06 (0.81) & 0.03 (0.81) & 0.04 (0.81) & 0.01 (0.81) & 0.1 (0.73) & 0.04 (0.73) & 0.09 (0.73) & 0.06 (0.73) & 0.04 (0.73) & 1.51 (1.42) \\ \hline
		
		LibreSSL & 1.87 (1.60) & 1.15 (2.07) & 0.62 (2.92) & 2.04 (1.60) & 2.99 (1.60) & {4.18} (1.60) & 2.39 (1.60) & 2.23 (1.04) & 1.92 (1.04) & 1.53 (1.04) & 1.14 (1.04) & 1.09 (1.04) & 4.18 (1.60) \\ \hline
		
		libsecp256k1 & -0.13 (0.08) & -0.27 (0.13) & 0.05 (0.21) & {0.21} (0.08) & 0.19 (0.08) & 0.14 (0.08) & 0.15 (0.08) & 0.07 (0.07) & -0.04 (0.07) & -0.19 (0.07) & 0.04 (0.07) & 0.02 (0.07) & 0.21 (0.08) \\ \hline
		
		libtomcrypt & 3.33 (1.28) & 2.42 (1.71) & 2.63 (2.39) & 3.06 (1.28) & 3.3 (1.28) & 2.93 (1.28) & 2.72 (1.28) & 2.83 (1.19) & 2.94 (1.19) & {3.43} (1.19) & 2.95 (1.19) & 2.36 (1.18) & 3.43 (1.19) \\ \hline
		
		mbedTLS & 0.84 (1.87) & 0.81 (2.46) & -2.2 (3.44) & 0.76 (1.87) & {0.85} (1.87) & -0.51 (1.88) & -0.65 (1.88) & 0.37 (1.68) & 0.49 (1.68) & 0.4 (1.68) & -0.17 (1.68) & -0.61 (1.69) & 0.85 (1.87) \\ \hline
		
		Monocypher & 0.38 (0.91) & 0.56 (1.45) & 0.2 (2.06) & -0.04 (0.91) & 0.28 (0.91) & {2.46} (0.91) & 1.65 (0.91) & 0.14 (0.87) & 0.0 (0.87) & 0.15 (0.87) & 0.77 (0.87) & 0.54 (0.87) & 2.46 (0.91) \\ \hline
		
		MS SymCrypt & {-0.33} (0.35) & -0.33 (0.35) & -0.31 (0.35) & -1.04 (0.35) & -0.34 (0.35) & -0.24 (0.35) & -0.33 (0.35) & -0.5 (0.31) & -0.08 (0.31) & -0.87 (0.31) & -0.72 (0.31) & -1.28 (0.31) & -0.33 (0.35) \\ \hline
		
		Nettle & 0.58 (1.0) & 0.78 (1.48) & 0.8 (2.53) & 0.61 (1.0) & 0.66 (1.0) & 0.61 (1.0) & 0.51 (1.0) & 0.83 (0.82) & {0.84} (0.82) & 0.73 (0.82) & 0.8 (0.82) & 0.66 (0.82) & 0.84 (0.82) \\ \hline
		
		OpenSSL & 2.28 (1.23) & 1.43 (1.58) & 1.23 (2.3) & 1.92 (1.23) & 2.01 (1.23) & 1.69 (1.23) & 1.73 (1.23) & 2.18 (1.09) & 2.39 (1.09) & {2.55} (1.09) & 1.83 (1.09) & 1.95 (1.09) & 2.55 (1.09) \\ \hline
		
		PQCrypto & -0.32 (0.56) & -0.07 (1.79) & 0.12 (2.83) & -0.14 (0.56) & 0.05 (0.56) & 0.0 (0.56) & -0.09 (0.56) & -0.03 (0.52) & -0.1 (0.52) & {0.14} (0.52) & 0.07 (0.52) & 0.01 (0.52) & 0.14 (0.52) \\ \hline
		
		SPHINCS+ & 0.01 (0.61) & {0.03} (1.01) & 0.03 (1.68) & 0.02 (0.61) & -0.02 (0.61) & 0.0 (0.61) & -0.01 (0.61) & 0.0 (0.59) & 0.0 (0.59) & -0.02 (0.59) & -0.01 (0.59) & 0.0 (0.59) & 0.03 (1.01) \\ \hline
		
		WolfSSL & 3.01 (1.96) & 2.45 (3.03) & 2.16 (4.95) & 3.25 (1.99) & {3.28} (1.99) & 2.52 (1.99) & 2.69 (1.99) & 2.92 (1.75) & 3.07 (1.75) & 2.99 (1.75) & 2.6 (1.75) & 2.64 (1.75) & 3.28 (1.99) \\ \hline
	\end{tabular}
}
\end{table}

\begin{table*}[tbp]
    \centering
    \caption{Performance speedups for cryptographic libraries with build-time mitigations, and code size \linebreak overheads in brackets. Results are shown across various aligned jump types, prefix sizes, and loop alignment boundaries, shown relative (\%) to each library’s default build. Negative values indicate slowdowns. The grey column indicates Intel’s default mitigation, and the bolded value marks the maximum speedup for each library. Averages are shown in the last row. 
    } 
    \label{tab:performance_impact_full}
    \scalebox{0.91}{
    \footnotesize
    \begin{tabular}{|l|c|c|c|c|c|c|c|c|c|c|c|c|}
        \hline
        \textbf{Branch} &\multicolumn{7}{c|}{ \cellcolor{gainsboro} \textbf{\texttt{jcc+fused+jmp}}} & \multicolumn{5}{c|}{\textbf{\texttt{jcc+fused}}} \\ \hline
        
        \textbf{Prefix} &  \multicolumn{3}{c|}{\cellcolor{gainsboro}\textbf{5}} & \textbf{4} & \textbf{3} & \textbf{2} & \textbf{1} & \textbf{5} & \textbf{4} & \textbf{3} & \textbf{2} & \textbf{1} \\ \hline
        
        \textbf{Loop} & \cellcolor{gainsboro}\textbf{16} & \textbf{32} & \textbf{64} & \multicolumn{9}{c|}{\textbf{16}} \\ \hline
        
        BearSSL & \cellcolor{gainsboro}9.62 (1.72) & 9.69 (1.72) & 9.49 (1.72) & 9.03 (1.72) & 9.73 (1.72) & 9.74 (1.72) & 9.29 (1.72) & 10.29 (1.60) & 10.06 (1.60) & 10.17 (1.60) & \textbf{10.54} (1.60) & 10.15 (1.60) \\ \hline
        
        BoringSSL & \cellcolor{gainsboro}0.83 (1.62) & 0.93 (2.31) & 1.16 (3.41) & 0.79 (1.62) & 0.78 (1.62) & 0.95 (1.62) & \textbf{1.23} (1.62) & 0.34 (1.37) & 0.51 (1.37) & 0.26 (1.37) & 0.29 (1.37) & 0.58 (1.37) \\ \hline
        
        Botan & \cellcolor{gainsboro}1.31 (1.55) & 1.27 (2.19) & -0.93 (3.37) & 0.18 (1.55) & 0.31 (1.55) & -1.34 (1.55) & -0.68 (1.55) & \textbf{1.74} (1.28) & -0.06 (1.28) & 0.16 (1.28) & 0.61 (1.28) & -0.97 (1.28) \\ \hline
        
        Crypto++ & \cellcolor{gainsboro}0.15 (0.83) & 0.16 (1.19) & 0.16 (1.76) & 0.11 (0.83) & \textbf{0.18} (0.83) & 0.11 (0.83) & 0.1 (0.83) & 0.08 (0.7) & 0.06 (0.7) & 0.02 (0.7) & -0.05 (0.7) & 0.07 (0.7) \\ \hline
        
        GnuTLS  & \cellcolor{gainsboro}0.25 (0.91) & 0.14 (0.91) & 1.85 (0.91) & \textbf{3.75} (0.91) & 2.97 (0.91) & 2.2 (0.91) & 0.98 (0.9) & 1.84 (0.71) & -0.09 (0.71) & -1.12 (0.71) & -0.96 (0.71) & -0.32 (0.71) \\ \hline
        
        libgcrypt & \cellcolor{gainsboro}0.04 (0.81) & \textbf{1.51} (1.42) & -0.01 (2.59) & 0.06 (0.81) & 0.03 (0.81) & 0.04 (0.81) & 0.01 (0.81) & 0.1 (0.73) & 0.04 (0.73) & 0.09 (0.73) & 0.06 (0.73) & 0.04 (0.73) \\ \hline
        
        LibreSSL & \cellcolor{gainsboro}1.87 (1.60) & 1.15 (2.07) & 0.62 (2.92) & 2.04 (1.60) & 2.99 (1.60) & \textbf{4.18} (1.60) & 2.39 (1.60) & 2.23 (1.04) & 1.92 (1.04) & 1.53 (1.04) & 1.14 (1.04) & 1.09 (1.04) \\ \hline
        
        libsecp256k1 & \cellcolor{gainsboro}-0.13 (0.08) & -0.27 (0.13) & 0.05 (0.21) & \textbf{0.21} (0.08) & 0.19 (0.08) & 0.14 (0.08) & 0.15 (0.08) & 0.07 (0.07) & -0.04 (0.07) & -0.19 (0.07) & 0.04 (0.07) & 0.02 (0.07) \\ \hline
        
        libtomcrypt & \cellcolor{gainsboro}3.33 (1.28) & 2.42 (1.71) & 2.63 (2.39) & 3.06 (1.28) & 3.3 (1.28) & 2.93 (1.28) & 2.72 (1.28) & 2.83 (1.19) & 2.94 (1.19) & \textbf{3.43} (1.19) & 2.95 (1.19) & 2.36 (1.18) \\ \hline
        
        mbedTLS & \cellcolor{gainsboro}0.84 (1.87) & 0.81 (2.46) & -2.2 (3.44) & 0.76 (1.87) & \textbf{0.85} (1.87) & -0.51 (1.88) & -0.65 (1.88) & 0.37 (1.68) & 0.49 (1.68) & 0.4 (1.68) & -0.17 (1.68) & -0.61 (1.69) \\ \hline
        
        Monocypher & \cellcolor{gainsboro}0.38 (0.91) & 0.56 (1.45) & 0.2 (2.06) & -0.04 (0.91) & 0.28 (0.91) & \textbf{2.46} (0.91) & 1.65 (0.91) & 0.14 (0.87) & 0.0 (0.87) & 0.15 (0.87) & 0.77 (0.87) & 0.54 (0.87) \\ \hline
        
        MS SymCrypt & \cellcolor{gainsboro}\textbf{-0.33} (0.35) & -0.33 (0.35) & -0.31 (0.35) & -1.04 (0.35) & -0.34 (0.35) & -0.24 (0.35) & -0.33 (0.35) & -0.5 (0.31) & -0.08 (0.31) & -0.87 (0.31) & -0.72 (0.31) & -1.28 (0.31) \\ \hline
        
        Nettle & \cellcolor{gainsboro}0.58 (1.0) & 0.78 (1.48) & 0.8 (2.53) & 0.61 (1.0) & 0.66 (1.0) & 0.61 (1.0) & 0.51 (1.0) & 0.83 (0.82) & \textbf{0.84} (0.82) & 0.73 (0.82) & 0.8 (0.82) & 0.66 (0.82) \\ \hline

        OpenSSL & \cellcolor{gainsboro}2.28 (1.23) & 1.43 (1.58) & 1.23 (2.3) & 1.92 (1.23) & 2.01 (1.23) & 1.69 (1.23) & 1.73 (1.23) & 2.18 (1.09) & 2.39 (1.09) & \textbf{2.55} (1.09) & 1.83 (1.09) & 1.95 (1.09) \\ \hline
        
        PQCrypto & \cellcolor{gainsboro}-0.32 (0.56) & -0.07 (1.79) & 0.12 (2.83) & -0.14 (0.56) & 0.05 (0.56) & 0.0 (0.56) & -0.09 (0.56) & -0.03 (0.52) & -0.1 (0.52) & \textbf{0.14} (0.52) & 0.07 (0.52) & 0.01 (0.52) \\ \hline
        
        SPHINCS+ & \cellcolor{gainsboro}0.01 (0.61) & \textbf{0.03} (1.01) & 0.03 (1.68) & 0.02 (0.61) & -0.02 (0.61) & 0.0 (0.61) & -0.01 (0.61) & 0.0 (0.59) & 0.0 (0.59) & -0.02 (0.59) & -0.01 (0.59) & 0.0 (0.59) \\ \hline

        WolfSSL & \cellcolor{gainsboro}3.01 (1.96) & 2.45 (3.03) & 2.16 (4.95) & 3.25 (1.99) & \textbf{3.28} (1.99) & 2.52 (1.99) & 2.69 (1.99) & 2.92 (1.75) & 3.07 (1.75) & 2.99 (1.75) & 2.6 (1.75) & 2.64 (1.75) \\ \hline
        
        \textbf{average} & \cellcolor{gainsboro}1.40 (1.11) & 1.33 (1.58) & 1.00 (2.32) & 1.45 (1.11) & \textbf{1.60} (1.11) & 1.50 (1.11) & 1.28 (1.11) & 1.50 (0.96) & 1.29 (0.96) & 1.20 (0.96) & 1.16 (0.96) & 1.00 (0.96) \\ \hline
    \end{tabular}
    }
\end{table*}

While this mitigation enables fusion for \obsA{} jumps and allows \obsB{} jumps to be cached in the \uop{} cache, the subsequent cache set still needs to be accessed. We hypothesize that aligning loops to 32B or 64B boundaries may further improve performance by reducing the number of \uop{} cache sets required to cache loops (which frequently execute jumps). Hence, we also experimented with loop alignment options. Note that for \texttt{O2} and \texttt{O3} optimization, the default loop alignment is 16 bytes.

\Cref{tab:performance_impact} summarizes the results for selected configurations, while our full results are presented in \Cref{tab:performance_impact_full}. On average, Intel’s default mitigation yields a 1.4\% speedup with a 1.11\% code size overhead across 17 cryptographic libraries. Specifically, 14 libraries show measurable speedups over their default builds. A notable case is BearSSL, which achieves a 9.62\% speedup at the cost of only a 1.72\% increase in binary size. In contrast, two libraries experience performance penalties of up to 0.33\%, which we attribute to interference with other low-level optimizations caused by the added padding. 

Our experiments (see Table~\ref{tab:performance_impact_full}) further show that loop alignment to 32 or 64 bytes does not consistently outperform Intel’s default configuration. Similarly, varying branch types and prefix sizes beyond the defaults does not reliably improve performance. However, two configurations stand out: Reducing the prefix size from 5 to 3 bytes (second column in \Cref{tab:performance_impact}) yields an average speedup of 1.6\%, with no code size increase over the default mitigation. Furthermore, disabling alignment for unconditional jumps (last column in \Cref{tab:performance_impact}) increases the speedup by 0.1\% while reducing the code size increase from 1.11\% to 0.96\%. { Finally, when selecting the configuration that maximizes speedup for each library, we observe an average speedup of 2.15\%---reaching up to 10.54\%---with a negligible average code size increase of 0.01\% over the default mitigation.}

These findings suggest that there remains untapped performance potential, not only within cryptographic libraries but also across widely used system libraries, {\emph{resulting in large-scale energy efficiency across millions of devices}. In particular, wolfSSL---with more than two billion daily connections worldwide in 2022~\cite{wolfssl2022report}---and Amazon s2n-tls---used across numerous AWS services---updated their libraries following our recommendations~\cite{wolfsslCommit151b9f0, s2ntlsV1_5_18}. Within} the Ubuntu 24.04.2 dataset, only a single library exhibited neither \obsA{} nor \obsB{} jumps—implying that 97\% of these libraries lack any mitigation against these slow-path cases and could benefit from adopting the proposed alignment strategy. Similarly, approximately 87\% of the analyzed Windows 10 Pro 22H2 libraries could be further optimized by ensuring that macro-fusible instruction pairs do not cross 32-byte boundaries.

\section{Covert Channels from Abnormal Conditional Jumps}
\label{sec:attack}

So far, we have analyzed how misaligned conditional jump instructions impact execution time due to the specific behavior of the x86 front-end. Previous work has shown that such timing variations can be exploited by an adversary to leak sensitive data by constructing a side or covert channel~\cite{aldaya2019portContention,ren2021deadUops,deng2022leakyFrontends}. We now quantify the extent to which these timing variations of misaligned conditional jump instructions can be exploited in a cross-core covert channel.

\subsection{Setup}

We consider the exemplary assembly code in Figure~\ref{lst:covert_channel} that performs conditional branching based on secret data, where one branch contains an ``abnormal'' jump affected by \obsA{} (line 23), while the other branch includes a ``normal'' jump (line 11). 
Based on our results in Section~\ref{sec:problem}, such an alignment should occur with a probability of 15\%. 

Our example code iterates over 256 secret bits, branching based on their value. To eliminate timing discrepancies due to different instructions in the two branches, we execute identical computations in both branches. This is common practice to eliminate timing side-channels, e.g., {when writing secure} cryptographic software. The \texttt{sub} and the conditional jump should be macro-fused unless the jump is aligned to a 64-byte boundary---this is the case for the jump in line 23---thereby adding a noticeable delay to the computation. 
As demonstrated in Section~\ref{sec:problem}, this alignment introduces measurable execution delays, enabling an adversary to infer secret-dependent {timing} variations.
We run this code on an Intel Xeon E-2176G CPU, operating at 4 GHz clock frequency, with 128GB RAM running Ubuntu 20.04 LTS.

\subsection{Covert-Channel}
We assume a sender who is running the code in Figure~\ref{lst:covert_channel}, while the receiver operates on a different physical core within the same machine. Moreover, we consider an example in which the sender and receiver synchronize both at the beginning and at the end of the code snippet shown in Figure~\ref{lst:covert_channel} using Intel's Transactional Synchronization Extensions (TSX). In line with prior work~\cite{DBLP:conf/eurosp/BriongosBMEM21}, the receiver gets the abort signal  immediately when the target code starts and ends, and uses these signals for measuring the execution time. 
Note that our covert channel does not depend on the availability of TSX, as synchronization can also be achieved through alternative mechanisms, including page faults~\cite{DBLP:conf/raid/JiangSK22/challengesDetectingSGX}, cache contention~\cite{DBLP:conf/raid/JiangSK22/challengesDetectingSGX,vanBulck2018foreshadowSGXoooExecution,DBLP:conf/hpca/HungerKRDVT15} or atomic operations on shared variables~\cite{DBLP:conf/ccs/BhattacharyyaSN19/smotherspectre}, as well as benign messages when permitted by the application~\cite{aldaya2019portContention}. These mechanisms can be used to signal the receiver when to start and stop execution-time measurements.

Our results are summarized in Figure~\ref{fig:covert_channel_errors}. We measure the error rates incurred by the different channel capacities.  More specifically, we encode the number of bits to be transmitted using the Hamming weight of the local variable accessed in line 17. For example, we can effectively send 1 bit from the sender to the receiver by setting the local variable in line 17 to have a Hamming weight of either 0 or 256. Similarly, the sender can encode up to 2 bits by setting the local variable to have a Hamming weight of 0 (transmitting bits ``00''), 83 (transmitting bits ``01''), 166 (transmitting bits ``10''), and 256 (transmitting bits ``11''), respectively (ensuring maximum spread of the Hamming weight of the local variable, hence maximizing the ability of the receiver to decode the message correctly). Notice that with a local variable size of 256 bits, the sender can encode at most 8 bits at a time on the channel; it is straightforward to see here that selecting a larger variable size allows the sender to encode additional bits per transmission. 

Our results suggest that one can leverage our covert-channel to encode up to 5 bits in approximately 1238.73 cycles with an accuracy as high as 95.46\%. In other words, we can achieve a maximum throughput of up to 16.14 Mbps (it takes on average 0.310 $\mu$s for the loop in Figure~\ref{lst:covert_channel} to complete in our setup) with an error rate of 4.54\%. In comparison, note that the maximum throughput achieved with the covert-channel in~\cite{wang2023crossCoreMacroOpFusion} is 858 kbps with a 5.85\% error rate. Similarly, the covert-channel  of~\cite{deng2022leakyFrontends} achieves a maximum throughput of 1.41 Mbps with 0\% error rate, and the one in~\cite{ren2021deadUops} achieves a throughput of 966 kbps with an error rate of 0.22\%.

\begin{table*}[tbp]
	\centering
	\small
	\caption{List of hardware performance counters (HPCs) on Intel CPUs and the specific events they capture.}
	\label{tab:hpc_list}
	\scalebox{0.98}{
		\begin{tabular}{|p{6.0cm}|l|} 
			\hline
			\textbf{HPC} & \textbf{Description} \\
			\hline
			\texttt{FRONTEND\_RETIRED.DSB\_MISS} & Number of retired instructions that experienced a \uop{} cache miss.\\ \hline
			\texttt{IDQ\_UOPS\_NOT\_DELIVERED.CYCLES\_0\_DELIV\_CORE} & Number of cycles the IDQ did not deliver any \uops{} to the back-end without back-end stalls. \\ \hline
			\texttt{UOPS\_ISSUED.STALL\_CYCLES} & Number of cycles the RAT did not issue any \uops{} to the RS. \\ \hline
			\texttt{UOPS\_ISSUED.ALL} & Number of \uops{} the RAT issued to the RS. \\ \hline
			\texttt{INSTRUCTIONS} & Number of instructions. \\ \hline
			\texttt{BACLEARS.ANY} & Number of front-end resteers due to a BPU misprediction. \\ \hline
	\end{tabular}}
\end{table*}

\begin{figure}[tbp]
	\centering
	\scriptsize
	\begin{Verbatim}[commandchars=\\\{\},frame=single,numbers=left,xleftmargin=5mm]
131a: lfence 
131d: rdtsc  
131f: mov    rdi,rax
1322: lfence 
1325: mov    eax,0xff
132a: mov    esi,0x1
132f: mov    r9d,0x1
1335: jmp    1340 <main+0x160>
1337: imul   r9,r12
\textcolor{blue}{133b: sub    eax,0x1}
\textcolor{blue}{133e: jb     1362 <main+0x182>}
1340: mov    edx,eax
1342: mov    ecx,eax
1344: sar    edx,0x6
1347: not    ecx
1349: movsxd rdx,edx
134c: mov    rdx,QWORD PTR [rsp+rdx*8+0x10]
1351: shl    rdx,cl
1354: test   rdx,rdx
1357: jns    1337 <main+0x157>
1359: imul   rsi,r12
\textcolor{red}{135d: sub    eax,0x1}
\textcolor{red}{1360: jae    1340 <main+0x160>}
1362: lfence 
1365: rdtsc 
	\end{Verbatim}
	\caption{Example of a loop that transmits information depending on branch location. The ``fast'' branch is highlighted in blue, and the ``slow'' branch in red.}
	\label{lst:covert_channel}
\end{figure}

\section{Conclusion}

In this paper, we measured and analyzed the timing variations of conditional jump instructions caused by their placement in a binary. Specifically, we demonstrate that the position of a conditional jump instruction affects macro-op fusion and the caching of the corresponding \uop{} in the \uop{} cache, which in turn influences instruction fetch time.
Through extensive measurements, we assessed the prevalence of this timing variability and found that up to 15.6\% of all conditional jumps incur additional execution cycles. We validate our analysis through a large-scale study of over 600 libraries in Windows 10 Pro 22H2 and Ubuntu 24.04.2, along with several widely used cryptographic libraries. Our results are striking: only one out of 320 Ubuntu libraries exhibited neither \obsA{} nor \obsB{} jumps; only 42 out of 323 Windows Pro libraries were free of jumps subject to \obsA{} or \obsB{}; and 0 of the 23 cryptographic libraries were not affected by \obsA{} nor \obsB.
We quantified the impact of this observation based on the cryptographic libraries,  showing an average speedup of 2.15\% and up to 10.54\% speedup when using the suggested mitigation.

Moreover, we show that these timing variabilities enable the creation of a practical covert channel, resulting in a throughput of up to 16.14 Mbps.
We therefore urge developers and practitioners to ensure that macro-fusible instructions are 32-byte aligned in their code. 
To ensure full reproducibility of our results, all associated source code and obtained measurements are available open-source at~\cite{our_artifact}.

\begin{figure}[t]
	\centering
	\includegraphics[width=0.93\columnwidth]{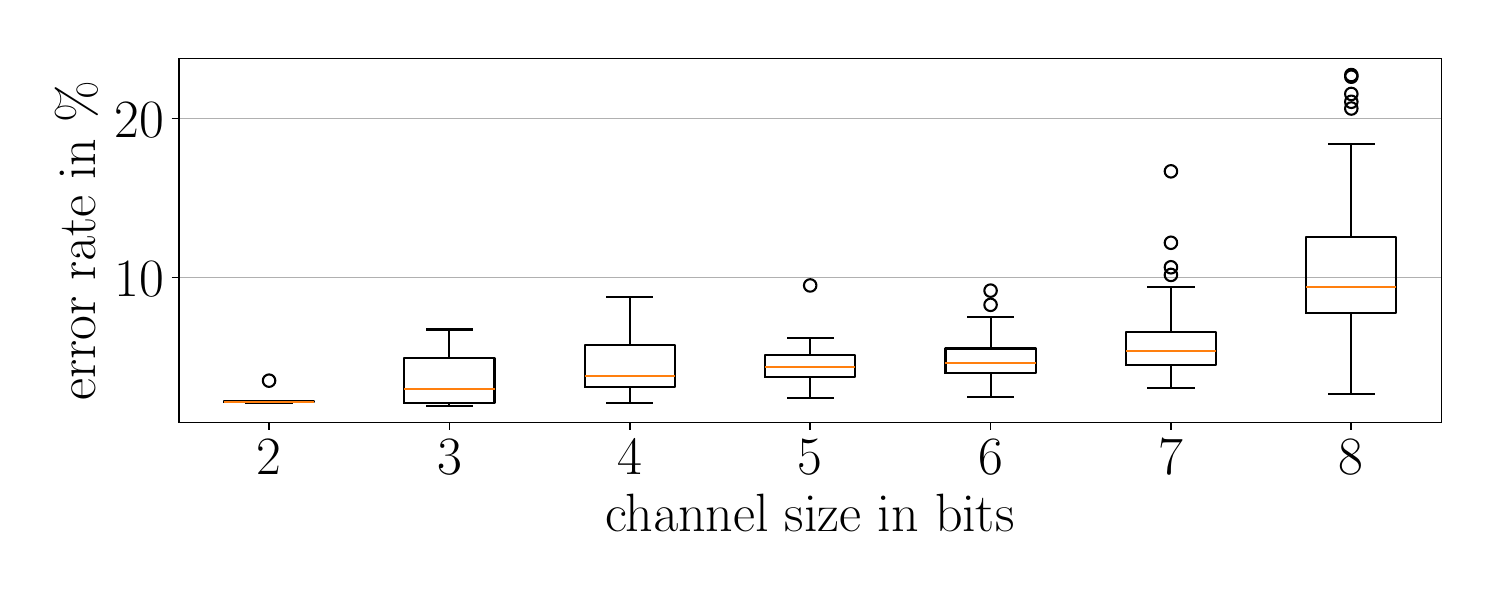}
	\caption{Error rates for different channel capacities based on our covert-channel. Here, every data point is averaged over 100 independent runs.}
	\label{fig:covert_channel_errors}
\end{figure}

\begin{acks}
This work is partly funded by the Deutsche Forschungsgemeinschaft (DFG, German Research Foundation) under Germany’s Excellence Strategy - EXC 2092 CASA - 390781972, and the European Union’s Horizon 2020 research and innovation program (REWIRE, Grant Agreement No. 101070627, and ACROSS, Grant Agreement No. 101097122). 
Views and opinions expressed are, however, those of the authors only and do not necessarily reflect those of the European Union. Neither the European Union nor the granting authority can be held responsible for them.  
\end{acks}

\bibliographystyle{ACM-Reference-Format}
\balance
\bibliography{references}

\end{document}